\documentclass[]{iopart}
\usepackage{natbib}

\usepackage[pdftex]{graphicx}
\usepackage{moreverb}
\usepackage[utf8]{inputenc}

\newcommand {\apgt} {\ {\raise-.5ex\hbox{$\buildrel>\over\sim$}}\ }
\newcommand {\aplt} {\ {\raise-.5ex\hbox{$\buildrel<\over\sim$}}\ }

\begin{document}

\title{High Performance Gravitational $N$-body Simulations on a
  Planet-wide Distributed Supercomputer}

\author{Derek Groen$^1$, Simon Portegies Zwart$^1$, Tomoaki Ishiyama$^2$, Jun Makino$^2$}

\address{$^1$ Leiden Observatory, Leiden University, P.O. Box 9513, 2300 RA Leiden, The Netherlands}
\address{$^2$ National Astronomical Observatory, Mitaka, Tokyo 181-8588, Japan}
\ead{djgroen@strw.leidenuniv.nl}

\maketitle

\begin{abstract}
We report on the performance of our cold-dark matter cosmological
$N$-body simulation which was carried out concurrently using
supercomputers across the globe.  We ran simulations on 60
to 750 cores distributed over a variety of supercomputers in Amsterdam
(the Netherlands, Europe), in Tokyo (Japan, Asia), Edinburgh (UK,
Europe) and Espoo (Finland, Europe).  Regardless the network latency
of 0.32 seconds and the communication over 30.000 km of optical
network cable we are able to achieve $\sim 87$\% of the performance
compared to an equal number of cores on a single supercomputer.  We
argue that using widely distributed supercomputers in order to acquire
more compute power is technically feasible, and that the largest
obstacle is introduced by local scheduling and reservation policies.
\end{abstract}

\section{Introduction}

Some applications for large scale simulations require a large amount of compute
power. This is often hard to acquire on a single machine. Combining multiple
supercomputers to do one large calculation can lift this limitation, but such
wide area computing is only suitable for certain algorithms. And even then the
political issues, like arranging the network, acquiring the compute time,
making reservations, scheduling runtime and synchronizing the run start, and
technical limitations are profound. Earlier attempts based on interconnecting
PC clusters were quite successful
\cite{inteugrid,Gualandris2007,Manos,Bal20083,QCGEscience2009}, but lacked the 
raw supercomputer performance required for our application.

Running simulations across multiple supercomputers has been done a few times
before \cite{IWay,paragon,arthropod,CosmoGrid}, though the performance
of simulations across three or more supercomputers has not yet been measured in
detail. Here we report on the performance of our parallel astronomical
simulations which use up to 4 supercomputers and predict the performance for
simulations which use 5 or more supercomputers.

In our experiments we use an international infrastructure of supercomputers.
These machines include an IBM Power6 supercomputer located at SARA in Amsterdam
(the Netherlands) and three Cray XT-4 supercomputers located at the Edinburgh
Parallel Computing Centre in Edinburgh (United Kingdom), the IT Center for
Science in Espoo (Finland) and the Center For Computational Astrophysics in
Tokyo (Japan). The Edinburgh site is equipped with a 1 Gbps interface while the
other three sites are equipped with a 10 Gbps interface. We achieved a peak
performance of 0.610 TFLOP/s and a sustained performance of 0.375 TFLOP/s using
120 cores distributed over 4 sites.  To provide a comparison with the
international tests we also run the code over up to 5 sites on a Dutch grid of
PC clusters. Our wide area simulations are realized with the development of a
software environment for Simulating the Universe Structure formation on
Heterogeneous Infrastructures, or SUSHI for short.


\section{Overview of SUSHI}\label{SUSHIoverview}

Our code is based on the GreeM cosmological $N$-body integrator, which was
originally developed for special-purpose GRAPE hardware
\cite{2005PASJ...57..849Y}. The code integrates the equations of motion for
dark matter particles using a shared and adaptive time step scheme and a 
Tree/Particle-Mesh (TreePM) force calculation method \cite{1995ApJS...98..355X} which assumes
periodic boundary conditions. The short range force interactions are resolved
using a Barnes-Hut tree algorithm \cite{1986Natur.324..446B} while the long
range interactions are resolved using a Particle Mesh (PM) algorithm
\cite{1981csup.book.....H}. 

The tree integration method places particles in a three-dimensional sparse
octree structure, where each cell contains the center of mass and the mass
aggregate of the particles therein. The method then resolves long range force
interactions using particles and tree cells instead of using direct
particle-particle evaluation. The accuracy of the tree integration method can
be tuned by changing the opening angle ($\theta$), which determines how small
and distant a group of particles needs to be to use the approximate
particle-tree cell evaluation. A higher value for $\theta$ results in fewer
particle-particle evaluations, and a lower accuracy of the simulation. Particle
integration using a tree algorithm is more compute-intensive than integration
using a PM algorithm, but we speed up the calculations by a factor $\sim 8$
using vector-math code optimizations for both the x86 \cite{2006NewA...12..169N}
and Power6 architectures. The PM algorithm maps the particles to a grid of 
mesh cells and calculates the
gravitational potential using the FFTW Fast Fourier Transform \cite{fftw}. It
accurately calculates the forces of long distance interactions, but is less
accurate in computing forces over short distances, for which the tree algorithm
is applied instead. 

The code has been modified to allow simulations on massively parallel machines
\cite{Ishiyama09}, in which the code uses a recursive multi-section scheme
\cite{2004PASJ...56..521M} to divide the workload over the processes. The
workload is redistributed during each step of the simulation so that the 
force calculation time remains equal for all processes.

\subsection{Parallelization across supercomputers}

We have developed SUSHI to efficiently use multiple supercomputers for our
simulations. We coupled the TreePM code with the MPWide communication library \cite{MPWide}
and developed a cross-supercomputer parallelization scheme. Because the wide
area network has performance and topological characteristics that are
different from local networks, the communication scheme between sites is
different from the scheme used between nodes. When SUSHI is deployed across
sites, each site is connected to two neighboring sites to form a ring 
topology.

\subsubsection{Communication scheme.}

A simulation using SUSHI consists of four communication phases per step. During 
these phases the simulation:
\begin{enumerate}
\item Exchanges mesh densities.
\item Collects sample particles to determine the site boundaries,
\item Exchanges tree structures with neighboring sites.
\item Migrates particles between neighboring sites.
\end{enumerate}

When exchanging mesh densities, the mesh cells from all sites are aggregated to
obtain the global mesh density. This mesh density is then used to perform PM
integration. In the code we have used here the PM integration is still a serial
operation, though the time spent on PM integration in our experiments is only a
small fraction of the total runtime. However, to support larger meshes
efficiently we will introduce parallelized PM integration in a later version of
the code. The mesh densities are gathered using a ring communication over all
sites. The force calculation time and time step information of each site are
also  accumulated during this phase.

At each step the site boundaries are updated, based on the obtained force
calculation times and the current particle distribution. To gain information on
the particle distribution, the communication processes on each site gather
sample particles from all other processes. These sampled particles are then
gathered from all sites using a ring communication.

Before the tree force calculations can be performed, each site constructs a
local essential tree structure. This local essential tree is a set of particles
and tree cells which are used to compute the force exchanges, and partially
resides on neighboring sites. To obtain a complete local essential tree, each
site therefore requires the missing tree data from its neighbors. The
simulation gathers the tree data using one neighbor exchange for two site runs,
or two exchanges for runs across three or more sites.

After the force calculations have been performed, the simulation updates the
positions of all particles. At this point, some particles may be located
outside the site boundaries and need to be migrated to a neighboring site. This
communication requires one neighbor exchange for two site runs, or two
exchanges for runs across three or more sites.

\subsubsection{Domain decomposition.}

We have implemented a hierarchical decomposition scheme to distribute the
particles among supercomputers. This scheme uses a one-dimensional slab
decomposition to distribute the particles among the sites, and a recursive
multi-section scheme over three dimensions to distribute the particles among
the processes. Because the domain decomposition between sites is
one-dimensional, each supercomputer only exchanges particles and tree
structures with two other machines. The data exchanges between sites can therefore
be done efficiently in a ring topology. Most supercomputers are connected by
optical paths in either a ring or star topology, if they are connected at all.

The simulation adjusts the distribution of particles among supercomputers at
run-time, such that the force calculation time is kept equal on all sites. The
number of particles $N_{i}$ on a single site $i$ for a run performed over $s$
sites is therefore given by

\begin{eqnarray}
  N_{i} \sim \frac{N \left( t_{\rm calc,i} \right)^{-1}} {\sum_{i=0}^{i<s} \left( t_{\rm calc,i} \right)^{-1}} .
  \end{eqnarray}

The force calculation time on site $i$ during the previous step is given by $t_{\rm calc,i}$.
The load balancing algorithm can be suppressed by explicitly limiting the boundary
moving length per step.

\subsubsection{Implementation of communication routines.}

We use the MPWide communication library \cite{MPWide} to
perform wide area message passing within SUSHI. The implementation of the
communication routines has few external dependencies, which makes it easy to
install on different platforms. The wide area communications in MPWide are
performed using parallel TCP streams. In cases where the supercomputers can
only be indirectly connected, we use MPWide-based port forwarding programs on
the intermediate nodes to establish a communication path. During the
development of SUSHI, we found that it is not trivial to obtain optimal
communication performance between supercomputers. Therefore we added several
features that can be used to improve the communication performance. The
communication routines within SUSHI can be customized for individual paths
between supercomputers. Settings that can be adjusted for each connection
include the number of parallel TCP streams, the TCP buffer sizes and the size
of data packages that are written to or read from the sockets. To improve the
performance on some long distance networks, MPWide also supports software-based
packet pacing.

\subsubsection{Memory consumption.}

Simulations run using SUSHI require 60 bytes of memory per particle and 52 
bytes per tree node. As we use an $N_{\rm leaf}$ value of 10 (where $N_{\rm
leaf}$ is the number of particles where the interaction tree will not be
divided further), we have on average 0.75 tree nodes per particle, and
therefore require a total of 99 bytes of memory per integrated particle (see
\cite{Ishiyama09} for further details). Consequently, a simulation with 
$N=2048^3$ requires 850 GB of RAM for tree integration while a run with
$N=8192^3$ requires at least 54 TB of RAM. In addition, 4.5 bytes per mesh cell
is required to do PM integration. These memory constraints place a lower limit
on the number of processes that can be used, and indirectly determine the
memory required on communication nodes. For particularly large exchange
volumes, the code can conserve memory on the communication nodes by 
communicating in multiple steps.

\section{Performance model}\label{SUSHIperfmodel}

We have developed a performance model for SUSHI. The model can be applied to
predict the execution time and scalability of simulations that run across
supercomputers. To make an accurate prediction we require several
architecture-dependent parameters. These include machine-specific parameters
such as the time spent on a single tree interaction (given by $\tau_{\rm
tree}$), the time required for one FFT operation $\tau_{\rm fft}$ and the time
required for the mesh operations on a single particle ($\tau_{\rm mesh}$). In
addition, we need a few parameters for the networks used in the simulation,
which are the round-trip time (given by $\lambda_{\rm lan}$ and $\lambda_{\rm
wan}$) and the available bandwidth ($\sigma_{\rm lan}$ and $\sigma_{\rm wan}$).
The values of these parameters can be obtained through minor network tests and
a small single process test simulation on each site. The values used for our
experiments are found in Tab.~\ref{Tab:Superspecs} for each supercomputer, in
Tab.~\ref{Tab:DAS3specs} for each national grid site and in
Tab.~\ref{Tab:PerfConstants} for the local and wide area networks.

\subsection{Single supercomputer}

The time required for one TreePM integration step using $p$ processes on a single
supercomputer ($t_{\rm exec}(1,p)$) consists of time spent on
tree force calculations ($t_{\rm tree}$), time spent on PM integration
($t_{\rm pm}$) and the communication overhead ($t_{\rm comm}$):

\begin{eqnarray}
  t_{\rm exec}\left( 1,p \right) &=& t_{\rm tree} + t_{\rm pm} + t_{\rm comm}.
\label{Eq:Texec}
\end{eqnarray}

The time spent on tree integration ($t_{\rm tree}$) is dominated by force
calculations. The force calculation time is obtained by multiplying the time
required to perform a single force interaction ($\tau_{\rm tree}$)
with the total number of tree interactions ($n_{\rm int}$) and
dividing it by the number of processes ($p$). Creating
interaction lists and constructing the tree introduce additional overhead
that scales with the number of interactions. To account for this in a simplified 
way, we multiply the time spent on force calculations
with a factor 1.2 \footnote{This value is based on timings from single site runs
using $N=128^3$ up to $N=2048^3$}. The time spent on tree integration then becomes

\begin{eqnarray}
  t_{\rm tree} &=& 1.2 \frac{\tau_{\rm tree} n_{\rm int}}{p}.
  \label{Eq:Ttree}
\end{eqnarray}

The number of interactions per simulation step depends on many parameters
including, but not limited to, the number of particles ($N$), the
opening angle of the tree integration ($\theta$) and the number of
mesh cells ($M$). We have performed several runs over a single
supercomputer and fitted the number of interactions for cosmological datasets,
which results in

\begin{eqnarray}
  n_{\rm int} = \frac{460 N^{1.0667}}{\theta^{1.35}} \frac{N^{1/12}}{M^{1/12} \sqrt{2.0}}.
\label{Eq:nInter}
\end{eqnarray}

In general, $n_{\rm int} \propto \theta^{-1.35}$ although this estimate may not
be accurate if $\theta \aplt 0.2$ or $\theta \apgt 0.75$. In these regimes, the
number of interactions depends more strongly on other tree integration
settings, such as the maximum number of particles allowed to share interaction
lists.

We calculate the time spent on PM integration ($t_{\rm pm}$) by adding the time
spent on the Fast Fourier Transform (FFT) to the time spent on mesh operations such as
interpolating the mesh values and setting the mesh forces on each particle. The
FFT requires $O(M \log_{2} M)$) FFT operations \cite{fftw}, each of which requires 
($\tau_{\rm fft}$) seconds. The time required for mesh operations scales with the 
local number of particles ($N/p$), where the time required per particle is equal to
the machine-specific constant $\tau_{\rm mesh}$. The total time spent on PM integration then 
becomes

\begin{eqnarray}
  t_{\rm pm} &=& \tau_{\rm fft} M \log_{2} M + \tau_{\rm mesh} \frac{N}{p} .
  \label{Eq:Tpm}
\end{eqnarray}

We decompose the communication time ($t_{\rm comm}$) into time spent to initially
traverse the networks ($t_{\rm l}$), which is latency-bound and time spent on
data throughput ($t_{\rm b}$), which is limited by the
available bandwidth of the local network. Therefore,

\begin{eqnarray}
 t_{\rm comm} &=& t_{\rm l} + t_{\rm b}.
\label{Eq:Tcomm}
\end{eqnarray}

For each step, the code performs 18 collective operations containing $\log p$
communication steps and two all-to-all communications with $p$ communication
steps. The time spent in latency ($t_{\rm l}$) is calculated by multiplying the
number of communication steps with the network round-trip time ($\lambda_{\rm
lan}$). As a result,

\begin{eqnarray}
 t_{\rm l} &=& \lambda_{\rm lan} \left( 18 \log p + 2 p \right),
\end{eqnarray}

We determine the time spent on data throughput ($t_{\rm b}$) by dividing the
data volume of the local area communications by the network bandwidth
($\sigma_{\rm lan}$). The communication data volume consists of three dominant
parts. These are the mesh cells residing at other processes (which is at most $4 M$ bytes in
total), the local essential tree structures (estimated to be $\left(48/\theta +
24\right) N^{2/3} p^{-2/3}$ bytes for $0.2 \aplt \theta \aplt 0.75$), and the 
sample particles which are used to determine the node boundaries ($12 N r_{\rm samp}$ bytes
in total). Here, $r_{\rm samp}$, which we set to $1/10000$ for large
calculations, is the ratio of sampled particles relative to $N$. The time spent
on data throughput is then 

\begin{eqnarray}
  t_{\rm b} &=& \frac{ 4 M + \left(144/\theta + 72\right) N^{2/3} p^{-2/3} + 12 N r_{\rm samp}}{\sigma_{\rm lan}}.
\end{eqnarray}

Additional communication is required to migrate particles between sites. The
data volume of this communication is relatively large during initial simulation
steps, but becomes negligible once sufficient steps have been taken to
adequately balance the workload. A detailed review of the communication
characteristics of the code is presented in \cite{Ishiyama09}.

\subsection{Multiple supercomputers}

We calculate the wall-clock time required for a single TreePM integration step
using $p$ processes in total across $s$ supercomputers ($t_{\rm exec}(s,p)$) by
adding the wide area communication overhead $w_{\rm comm}$ to the time spent on
tree integrations ($t_{\rm tree}$), the time spent on PM integration ($t_{\rm
pm}$) and the time spent local area communications ($t_{\rm comm}$). The 
execution time per step is therefore 

\begin{eqnarray}
  t_{\rm exec}\left( s,p \right) &=& t_{\rm tree} + t_{\rm pm} + t_{\rm comm} + w_{\rm comm}.
\end{eqnarray}

Here, we calculate $t_{\rm tree}$ using Eq.~\ref{Eq:Ttree}, $t_{\rm pm}$ using
Eq.~\ref{Eq:Tpm} and $t_{\rm comm}$ using Eq.~\ref{Eq:Tcomm}. Note that for
runs across sites we calculate the local latency-bound communication time
$t_{\rm l}$ using the number of local processes $p/s$, rather than $p$. The
communication overhead on the wide area network ($w_{\rm comm}$) consists of
the time spent in latency ($w_{\rm l}$) and the time spent on data throughput
($w_{\rm b}$) on the wide area network. As a result, 

\begin{eqnarray}
 w_{\rm comm} &=& w_{\rm l} + w_{\rm b}.
\end{eqnarray}

The code performs five blocking gather operations over all sites per step. These
gathers are performed using a ring scheme, which requires $s-1$ neighbor exchanges
per gather. We also require four blocking exchanges with each of the two 
neighboring sites. The total number of exchanges is then equal to $5s+3$ and,
the total time spent in latency ($w_{\rm l}$) then becomes

\begin{eqnarray}
 w_{\rm l}  &=& \lambda_{\rm wan} \left( 5 s + 3 \right).
\end{eqnarray}

Here, $\lambda_{\rm wan}$ is the network round-trip time between sites. 

We calculate the time spent on wide area data throughput ($w_{\rm b}$) by
dividing the data volume of the wide area communications by the bandwidth
capacity of the wide area network ($\sigma_{\rm wan}$). The volume of the
exchanged data between sites is similar to the data volume between nodes with
three exceptions. First, the exchange of mesh densities requires one float per
mesh cell per site. Second, because SUSHI uses a 1D decomposition between sites
the volume of the local essential tree is larger. Third, because of the 1D
decomposition we exchange only the Cartesian {\tt x} coordinates of sampled
particles. The data volume for the exchange of sampled particles is therefore
three times smaller. The total time spent on wide area data throughput is 

\begin{eqnarray}
 w_{\rm b} &=& \frac{4 s M + \left(48/\theta + 24\right) N^{2/3} + 4 N r_{\rm samp}}{\sigma_{\rm wan}}.
\end{eqnarray}

\subsection{Scalability across sites}

The speedup of a simulation across sites, $S \left( s \right)$, is defined by
dividing the time required for an integration step on 1 site using $p$ processes
($t_{\rm exec} \left( 1,p \right)$) by the time required for an integration step
over $s$ sites using a total of $s \cdot p$ processes ($t_{\rm exec}\left( s,
s p \right)$). It is therefore given by

\begin{eqnarray}
 S \left( s \right) = \frac{t_{\rm exec} \left( 1,p \right) }{t_{\rm exec} \left( s, s p \right) }.
\label{Eq:sp}
\end{eqnarray}

The efficiency of a simulation across sites, $E \left( s \right)$, is calculated
by dividing the time required for an integration step on 1 site using $p$
processes by the time required for an integration step over $s$ sites using a
total of $p$ processes ($t_{\rm exec}\left( s, p \right)$). The efficiency is
then

\begin{eqnarray}
 E \left( s \right) = \frac{t_{\rm exec} \left( 1,p \right) }{t_{\rm exec} \left( s, p \right) }.
\label{Eq:eff}
\end{eqnarray}

\section{Experiments}\label{SUSHIexperiments}

We have tested SUSHI for performance on a grid of 5 Beowulf clusters, as well
as an infrastructure consisting of four supercomputers. Each simulation lasts
for 100 integration steps, and uses an opening angle of $\theta=0.3$ when the 
redshift $z > 10$ and $\theta=0.5$ when $z \le 10$. Here the redshift $z$ is 
used as an indication of time, with the Big Bang occuring at $z = \infty$ and
$z = 0$ being the present day. For each opening angle we  measured the total 
wall-clock time and communication time per step, averaged over 10 steps. All 
measurements were made near $z=10$, approximately 460 million years after the 
Big Bang. A full listing of the simulation parameters and initial condition
characteristics of our experiments is given in Tab.~\ref{Tab:ICspecs}. 

\begin{table}
\centering
      \begin{tabular}{ll}
      \hline
      Parameter                                       & Value \\
      \hline
      Matter density parameter ($\omega_0$)           & 0.3 \\
      Cosmological constant ($\lambda_0$)             & 0.7 \\
      Hubble constant ($H_0$)                         & 70.0 km/s/Mpc \\
      Box size                                        & $(30 Mpc)^3$ \\
      Mass fluctuation parameter ($\sigma_8$)         & 0.9 \\
      Softening for $N=256^3$/$512^3$/$1024^3$ run    & 5/2.5/1.25 Kpc \\
      Sampling rate $r_{\rm samp}$ for $N=256^3$      & 2500 \\
      Sampling rate $r_{\rm samp}$ for $N > 256^3$    & 10000 \\
      Tree opening angle ($\theta$), $z > 10$         & 0.3 \\
      Tree opening angle ($\theta$), $z \le 10$       & 0.5 \\
      Tree ncrit                                      & 1000 \\
      \hline
    \end{tabular}

  \caption{Initial condition and accuracy parameters used for our test
  simulations. The maximum number of particles allowed to share an interactions
  list in the tree integration is given by ncrit.}
  
  \label{Tab:ICspecs}
\end{table}

We compare the results of our experiments with predictions from our performance
model. To do so, we measured the value of several machine constants using local
tests and provide them in Tab.~\ref{Tab:DAS3specs} and
Tab.~\ref{Tab:Superspecs}. The network constants used in the performance model
are given in Tab.~\ref{Tab:PerfConstants}. As all simulations lasted for 100 
integration steps, particle exchanges were still performed to improve the
distribution of work. We have added the measured average data volume of these
exchanges to the data volume for wide area communications in our model. For
full-length simulations, the data volume of these particle exchanges becomes a
negligible part of the total communication volume.

\subsection{DAS-3 experiment setup}\label{Sec:Das3expsetup}


The Distributed ASCI Supercomputer 3 (DAS-3 \cite{DAS3}) is a Dutch
infrastructure that consists of 5 PC clusters within The Netherlands. The
clusters use 10 Gbps networking internally, while the head nodes of each site
are connected to regular internet in a star
topology. Using end-to-end message passing tests we were able to achieve a
performance of up to 1.25 Gbps between sites. The specifications of the
five DAS-3 sites can found in Tab.~\ref{Tab:DAS3specs}.

\begin{table}
\centering
    \begin{tabular}{|l|l|l|l|l|l|}

\hline
Site                 & VU      & UvA     & LIACS   & TU      & MM \\
\hline
City                 & A'dam   & A'dam   & Leiden  & Delft   & A'dam \\

AMD CPU model        & DP280   & DP275   & DP252   & DP250   & DP250 \\

Number of nodes      & 85      & 41      & 32      & 68      & 46 \\

Cores per node       & 4       & 4       & 2       & 2       & 2 \\

CPU freq. [GHz]      & 2.4     & 2.2     & 2.6     & 2.4     & 2.4 \\

Memory / core [GB]   & 1       & 1       & 2       & 2       & 2 \\

Peak [TFLOP/s]       & 3.26    & 1.44    & 0.66    & 1.31    & 0.88 \\ 

$\tau_{\rm tree}$ [$\times 10^{-9}$ s] & 5.9 & 6.4 & 5.4  & 5.9  & 5.9\\

$\tau_{\rm fft}$  [$\times 10^{-9}$ s] & 5.0  & 5.0  & 5.7  & 5.0  & 5.0 \\

$\tau_{\rm mesh}$ [$\times 10^{-6}$ s] & 2.4  & 2.4  & 2.4  & 1.9  & 2.4 \\

Ordering             &$1^{st}$ &$2^{nd}$ &$3^{rd}$ &$4^{th}$ &$5^{th}$ \\

\hline

      \end{tabular}
  \caption{Technical specifications of five sites of the DAS-3 Dutch Grid. Three of the five DAS-3 sites reside in Amsterdam.}
  \label{Tab:DAS3specs}
\end{table}

We performed experiments using three problem sizes and two opening angles. For
our tests, we performed one set of runs with $N=256^3$ particles and $M=128^3$
mesh cells using 60 processes distributed evenly among the sites and two sets
of runs with $N=512^3$ particles using 120 processes in total. One of the runs
with $N=512^3$ uses $M=128^3$ and the other uses $M=256^3$. We maintained a
fixed site ordering for all our runs as given in the bottom row of
Tab.~\ref{Tab:DAS3specs}. For all our experiments on the DAS-3, we have configured
MPWide to use a single TCP stream per communication channel, send
messages in chunks of 8 kB and receive messages in chunks of 32 kB. We did
not use software-based packet pacing during these tests.

\subsection{DAS-3 results}

\begin{figure}
  \centering
    \includegraphics[width=0.8\columnwidth]{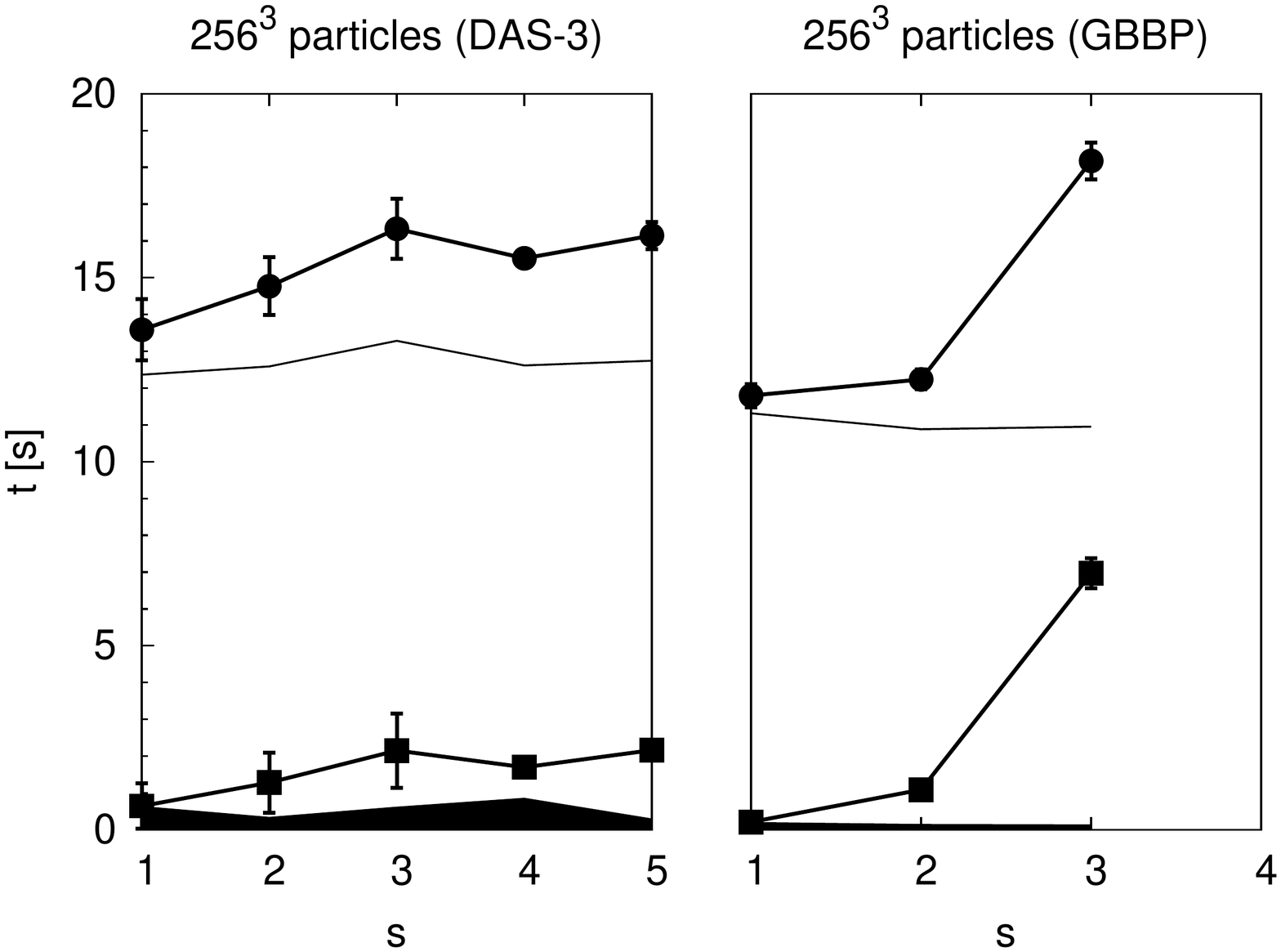}

      \caption{Wall-clock time (given by circles) and total communication time
(including local and wide area exchanges, given by squares) per step averaged
over 10 steps as a function of the number of sites $s$ for two different
simulations. We also included the average time spent on tree force calculations
(given by the thin lines without symbols), and the time spent on local area
communications (given by the filled surface area at the bottom). Results are
given for a simulation with $N=256^3$ and $M=128^3$ over 60 processes in total
using the DAS-3 (left panel) and across multiple supercomputers (right panel).
All the runs use $\theta=0.3$. The standard deviation of each timing measurement 
is shown by an error bar. The average time spent on PM integration was 1.07-1.41 s 
per step for the DAS-3 runs, and 0.42-0.45 s for the GBBP runs.}

  \label{Fig:256runs}
\end{figure}

\begin{figure}
  \centering
    \includegraphics[width=0.8\columnwidth]{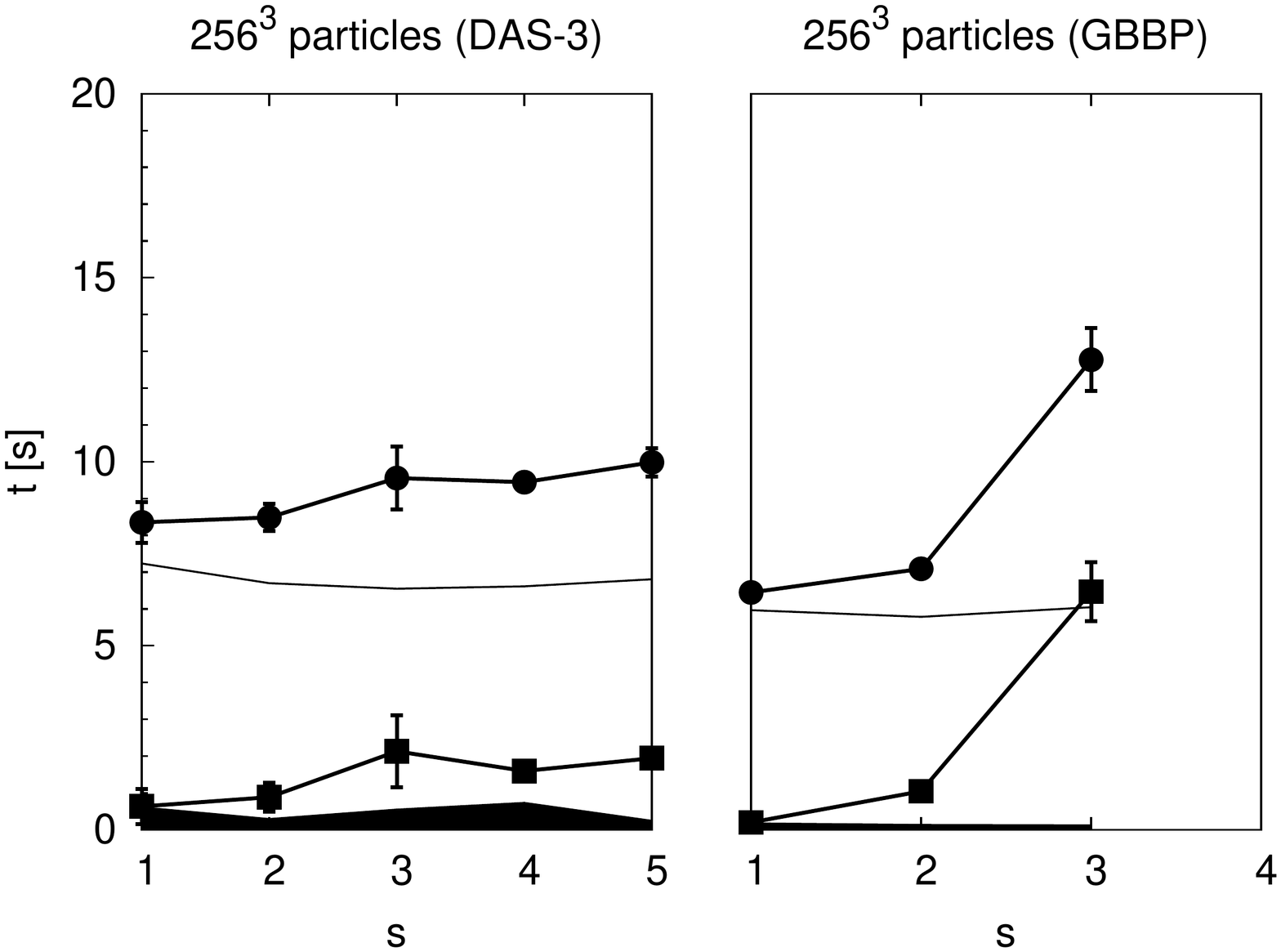}

      \caption{As Fig.~\ref{Fig:256runs}, but for simulations using $\theta=0.5$. 
      The average time spent on PM integration was 1.07-1.41 s per step for the 
      DAS-3 runs, and 0.42-0.45 s for the GBBP runs.}

  \label{Fig:256runs2}
\end{figure}

\begin{figure}
  \centering
    \includegraphics[width=0.8\columnwidth]{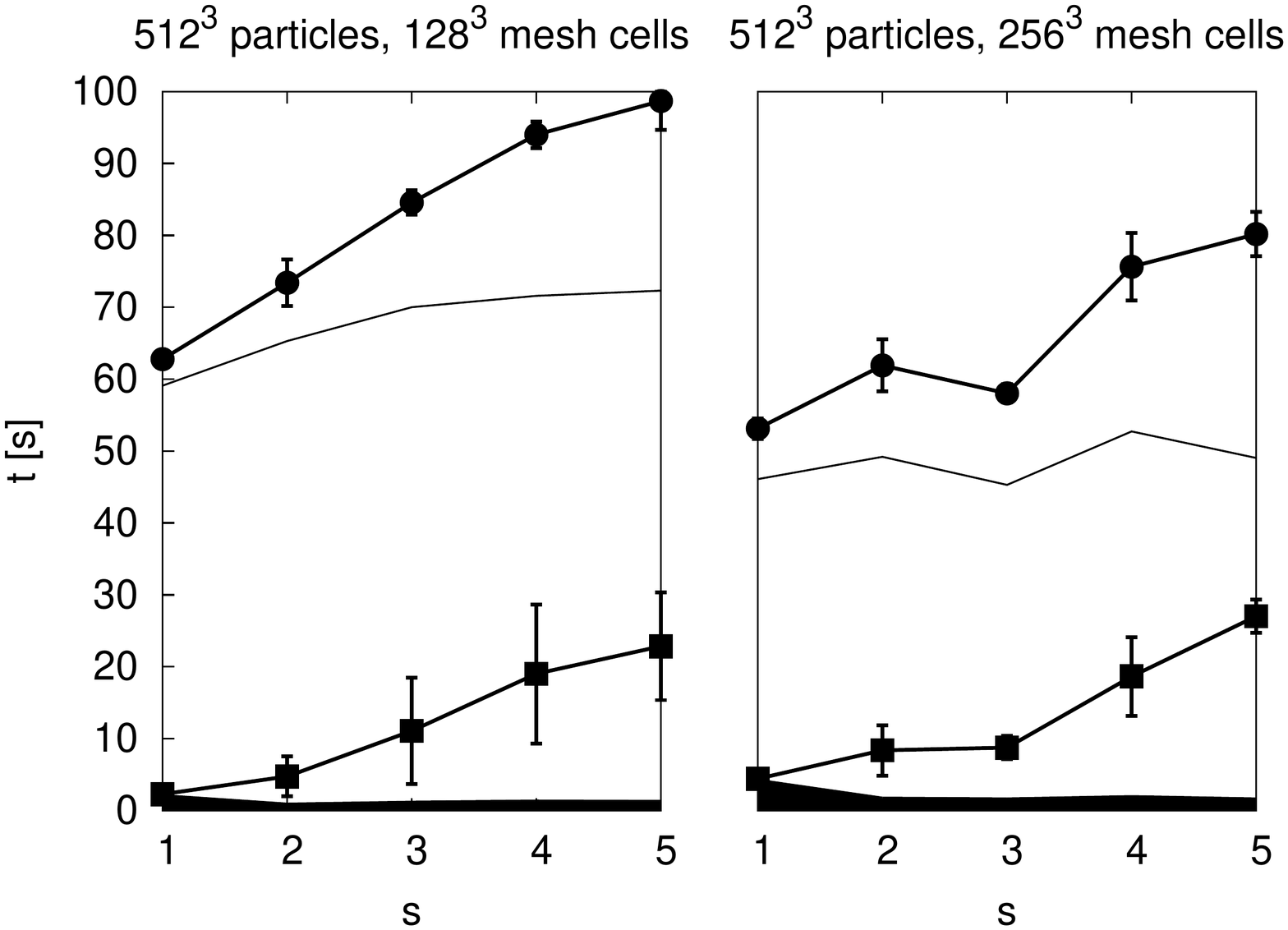}
      \caption{As Fig.~\ref{Fig:256runs}, but for simulations on the DAS-3 
      using $N=512^3$ and $M=128^3$ (left panel), and simulations using 
      $N=512^3$ and $M=256^3$ (right panel). All simulations were run over
      120 processes in total. The average time spent on PM integration was 
      3.16-3.57 s for the runs using $M=128^3$, and 5.77-6.48 s for the runs
      using $M=256^3$.}
  \label{Fig:Das3runs}
\end{figure}

\begin{figure}
  \centering
    \includegraphics[width=0.8\columnwidth]{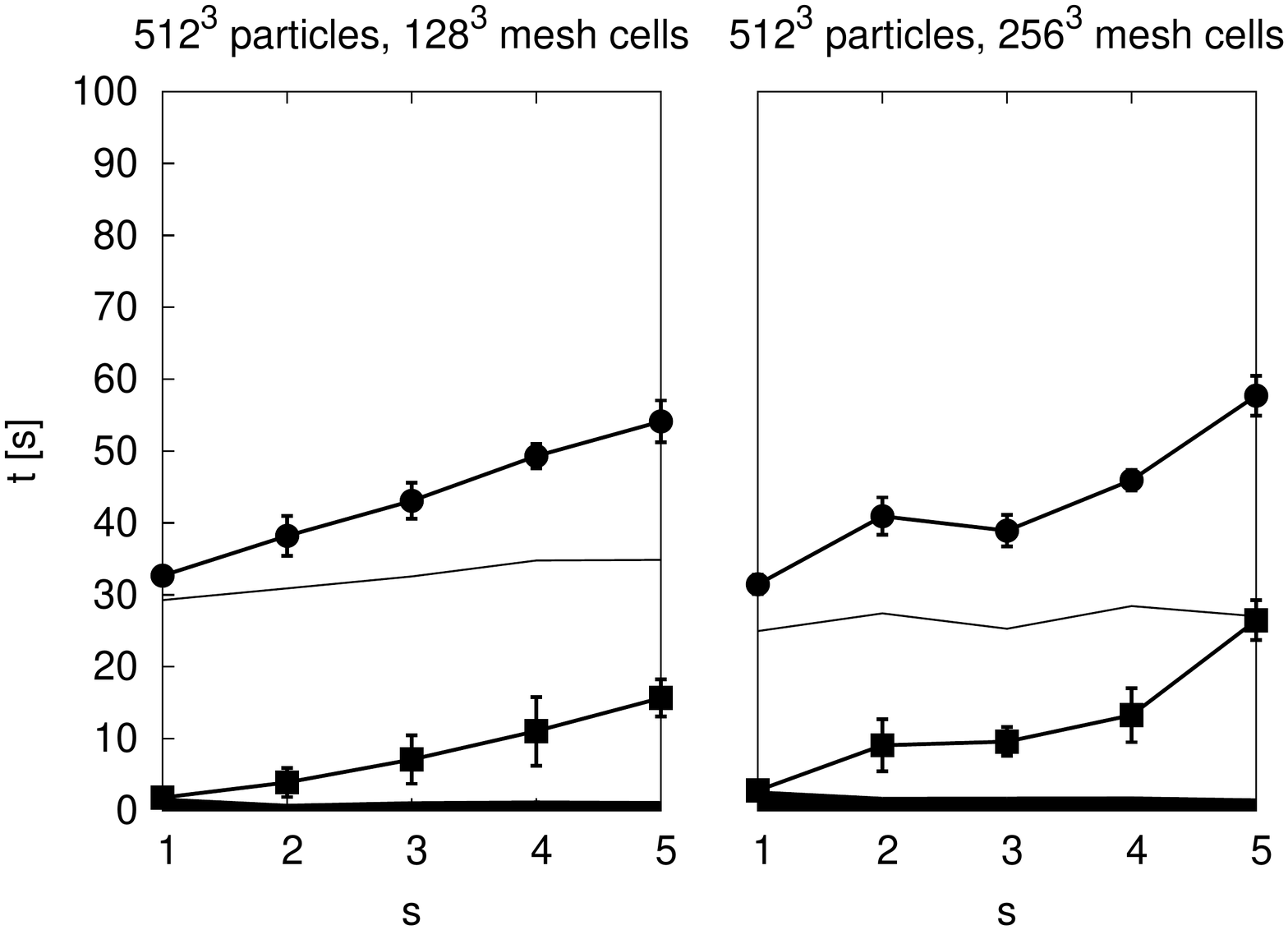}
      \caption{As Fig.~\ref{Fig:Das3runs}, but for simulations using 
      $\theta=0.5$. The average time spent on PM integration was
      3.16-3.57 s for the runs using $M=128^3$, and 5.77-6.48 s for the runs
      using $M=256^3$.}
  \label{Fig:Das3runs2}
\end{figure}

The timing measurements of our experiments can be found respectively in the
left panel of Fig.~\ref{Fig:256runs} and \ref{Fig:256runs2} for the runs with $N=256^3$ and in
Fig.~\ref{Fig:Das3runs} and \ref{Fig:Das3runs2} for the runs with $N=512^3$. Here we see that the
total communication overhead (including both local and wide area communications) becomes
marginally higher as we increase in number of sites in the simulation with
$N=256^3$ and $M=128^3$ mesh cells, both for runs with $\theta=0.3$ and with
$\theta=0.5$. The measured total communication overhead for simulations with
$N=512^3$ increases more steeply with $s$, because the larger communication
data volume results in a higher traffic load on the internet lines between the
sites. On the other hand, the time spent on local communications becomes lower 
when we run SUSHI across more sites. The number of processes per site ($p/s$) 
is lower for runs across more sites, which results in a smaller load and less
congestion on the local network.

The wall-clock time per step is almost twice as high for runs using
$\theta=0.3$ compared to runs with $\theta=0.5$, while the communication
overhead is approximately equal. The runs with $M=128^3$ have a total 
execution time which scales more steeply with $s$ than the
communication time. The run with $N=512^3$ and $M=128^3$ require
more time to achieve load balance than runs with $M=256^3$ due to 
the more compute-intensive tree integrations, which amplify the difference
in CPU frequencies between the DAS-3 sites. The time spent on tree
integration increases with $s$ in the runs with $M=128^3$, which indicates
that good load balance was not yet achieved here.

We provide a comparison between our timing results from the experiments and
predictions from our performance model in Tab.~\ref{Tab:Das3runs03} and
Tab.~\ref{Tab:Das3runs05}. Here we see that the achieved performance roughly
matches the model predictions, with the notable exception for the local 
communications. In our model we used $\sigma_{\rm lan} = 1 \times 10^8$, which
results in much lower communication times than those we have measured 
in our experiments. As the local communication performance is dominated by
bandwidth rather than latency, we conclude that the achieved intra-site
point-to-point bandwidth using local MPI has been considerably lower than 
$\sim 100 MB/s$.
 
For the runs with 
$N=512^3$, the model tends to underestimate the wall-clock and communication times 
for $M=128^3$ due to the earlier mentioned load balance issues in these 
runs. However, we find slightly higher times in the model for runs with 
$M=256^3$ than in the results. In our model the size of the local essential 
tree is only dependent on $N$ and $\theta$, but we also find a minor correlation 
with the number of mesh cells used in our experiment results, as the range of the 
tree integration is equal to three mesh cell lengths.


\begin{table}
\centering  
    \begin{tabular}{lllllllllllll}

\hline
$N^{1/3}$ & $M^{1/3}$ & np & $s$ & \multicolumn{2}{c}{comm.} & tree & exec. & \multicolumn{2}{c}{$t_{\rm comm}$} & $t_{\rm tree}$ & $t_{\rm exec}$ \\

     &     &     &        & local  & total &      &        & only  & $+w_{comm}$ &      &      \\
\hline
     &     &     &        & real   & real  & real & real   & model & model       & model & model\\
     &     &     & \#     & [s]    & [s]   & [s] & [s]     & [s]   & [s]         & [s]    & [s] \\    
\hline
256  & 128 & 60  & 1 & 0.644 & 0.644 & 12.36 & 13.59 & 0.13 & 0.13 & 11.79 & 12.81\\
256  & 128 & 60  & 2 & 0.346 & 1.277 & 12.59 & 14.77 & 0.12 & 0.73 & 12.29 & 13.91\\
256  & 128 & 60  & 3 & 0.628 & 2.152 & 13.29 & 16.33 & 0.12 & 1.63 & 11.79 & 14.35\\
256  & 128 & 60  & 4 & 0.865 & 1.695 & 12.61 & 15.52 & 0.11 & 2.85 & 11.79 & 15.53\\
256  & 128 & 60  & 5 & 0.304 & 2.160 & 12.74 & 16.15 & 0.11 & 4.40 & 11.79 & 17.09\\
\hline
512  & 128 & 120 & 1 & 2.275 & 2.275 & 59.09 & 62.79 & 0.18 & 0.18 & 64.44 & 67.52\\
512  & 128 & 120 & 2 & 1.092 & 4.745 & 65.29 & 73.41 & 0.16 & 2.84 & 67.17 & 72.92\\
512  & 128 & 120 & 3 & 1.376 & 11.07 & 70.01 & 84.56 & 0.16 & 5.82 & 64.44 & 73.19\\
512  & 128 & 120 & 4 & 1.523 & 19.01 & 71.59 & 94.01 & 0.15 & 9.11 & 64.44 & 76.35\\ 
512  & 128 & 120 & 5 & 1.471 & 22.84 & 72.31 & 98.68 & 0.15 & 12.73& 64.44 & 79.99\\
\hline
512  & 256 & 120 & 1 & 4.421 & 4.421 & 46.09 & 53.12 & 0.74 & 0.74 & 54.19 & 59.62\\
512  & 256 & 120 & 2 & 1.921 & 8.368 & 49.20 & 61.90 & 0.72 & 5.64 & 56.48 & 66.83\\
512  & 256 & 120 & 3 & 1.830 & 8.752 & 45.28 & 58.03 & 0.72 & 13.10& 54.19 & 72.26\\
512  & 256 & 120 & 4 & 2.137 & 18.65 & 52.72 & 75.64 & 0.71 & 23.11& 54.19 & 82.14\\
512  & 256 & 120 & 5 & 1.815 & 27.03 & 49.03 & 80.17 & 0.71 & 35.69& 54.19 & 94.74\\
\hline
      \end{tabular}

  \caption{List of runs performed on the DAS-3, which use $\theta=0.3$. The
cube root of the number of particles and mesh cells are given in the first and
second column, the number of processes and the number of sites involved in the
third and fourth column. The next four columns contain timing measurements from
our experiments, which are average times per step averaged over 10 steps. The
columns contain respectively the intra-site communication time, the total
communication time, the time spent on tree integration (excluding PM) and the
total wall-clock time. The last four columns contain the intra-site
communication time, total communication time, tree integration time (excluding PM), and
wall-clock time as predicted by our performance model.}

\label{Tab:Das3runs03}
\end{table}

\begin{table}
\centering 
    \begin{tabular}{lllllllllllll}

\hline
$N^{1/3}$ & $M^{1/3}$ & np & $s$ & \multicolumn{2}{c}{comm.} & tree & exec. & \multicolumn{2}{c}{$t_{\rm comm}$} & $t_{\rm tree}$ & $t_{\rm exec}$ \\

     &     &     &        & local  & total &      &        & only  & $+w_{comm}$ &      &      \\
\hline
     &     &     &        & real   & real  & real & real   & model & model       & model & model\\
     &     &     & \#     & [s]    & [s]   & [s] & [s]     & [s]   & [s]         & [s]    & [s] \\
\hline
256  & 128 & 60  & 1 & 0.631 & 0.631 & 7.233 & 8.351 & 0.12  & 0.12  & 5.92  & 6.93\\
256  & 128 & 60  & 2 & 0.314 & 0.884 & 6.698 & 8.488 & 0.11  & 0.64  & 6.17  & 7.70\\
256  & 128 & 60  & 3 & 0.571 & 2.131 & 6.547 & 9.555 & 0.11  & 1.46  & 5.92  & 8.30\\
256  & 128 & 60  & 4 & 0.757 & 1.593 & 6.615 & 9.446 & 0.11  & 2.61  & 5.92  & 9.41\\
256  & 128 & 60  & 5 & 0.269 & 1.944 & 6.806 & 9.981 & 0.10  & 4.07  & 5.92  & 10.88\\
\hline
512  & 128 & 120 & 1 & 1.797 & 1.797 & 29.26 & 32.67 & 0.16  & 0.16  & 32.33 & 35.40\\
512  & 128 & 120 & 2 & 0.947 & 3.899 & 30.89 & 38.20 & 0.14  & 2.50  & 33.70 & 39.11\\
512  & 128 & 120 & 3 & 1.286 & 7.076 & 32.55 & 43.08 & 0.14  & 5.16  & 32.33 & 40.43\\
512  & 128 & 120 & 4 & 1.392 & 11.05 & 34.77 & 49.30 & 0.13  & 8.13  & 32.33 & 43.26\\ 
512  & 128 & 120 & 5 & 1.318 & 15.65 & 34.86 & 54.11 & 0.13  & 11.43 & 32.33 & 46.58\\
\hline
512  & 256 & 120 & 1 & 2.781 & 2.781 & 24.95 & 31.47 & 0.72  & 0.72  & 27.19 & 32.60\\
512  & 256 & 120 & 2 & 1.897 & 9.067 & 27.43 & 40.93 & 0.70  & 5.30  & 28.34 & 38.34\\
512  & 256 & 120 & 3 & 1.923 & 9.586 & 25.29 & 38.94 & 0.70  & 12.44 & 27.19 & 44.61\\
512  & 256 & 120 & 4 & 1.955 & 13.27 & 28.45 & 45.97 & 0.69  & 22.13 & 27.19 & 54.16\\
512  & 256 & 120 & 5 & 1.694 & 26.47 & 27.01 & 57.70 & 0.69  & 34.39 & 27.19 & 66.44\\
\hline
      \end{tabular}

  \caption{As Tab.~\ref{Tab:Das3runs03}, but with $\theta=0.5$ instead
  of $\theta=0.3$.}

\label{Tab:Das3runs05}
\end{table}

\subsection{Gravitational Billion Body Project experiment setup}

We have run a number of test simulations across multiple supercomputers
to measure the performance of our code and to test the validity of our
performance model. The simulations, which use datasets consisting of $N=256^3$,
$N=512^3$ and $N=1024^3$ dark matter particles, were run across up to four
supercomputers. We provide the technical characteristics of each supercomputer
in Tab.~\ref{Tab:Superspecs}. The three European supercomputers are connected 
to the DEISA shared network, which can be used without prior
reservation although some user-space tuning is required to get acceptable
performance. The fourth supercomputer resides in Japan and is connected to the
other three machines with a 10~Gbps intercontinental light path. 

\subsubsection{Network configuration.}

In the shared DEISA network we applied the following settings to MPWide: First,
all communication paths used at least 16 parallel streams and messages were 
sent and received in chunks of 256~kB per stream. These settings allow
us to reach $\sim 100$~MB/s sustained throughput on the network between
Amsterdam and Edinburgh. Second, we used software-based packet pacing to reduce
the CPU usage of MPWide on the communication nodes. This had little impact on
the communication performance of the application, but was required because some
of the communication nodes were non-dedicated.


Although the light path between Amsterdam and Tokyo did not have an optimal TCP
configuration, we were able to to achieve a sustained throughput rate of $\sim$
100 MB/s by tuning our MPWide settings. To accomplish this throughput rate, we
used 64 parallel TCP streams, limited our burst exchange rate to 100 MB/s per
stream using packet pacing and performed send/receive operations in chunks
of 8kB per stream. In comparison, when using a single TCP stream, our
throughput was limited to 10~MB/s, even though the TCP buffering size was set
to more than 30~MB on the end nodes. We believe that this limitation arises
from TCP buffer limitations on one of the intermediary nodes on the light path.

\begin{figure}

  \begin{tabular}{cc}
  \includegraphics[width=0.47\columnwidth]{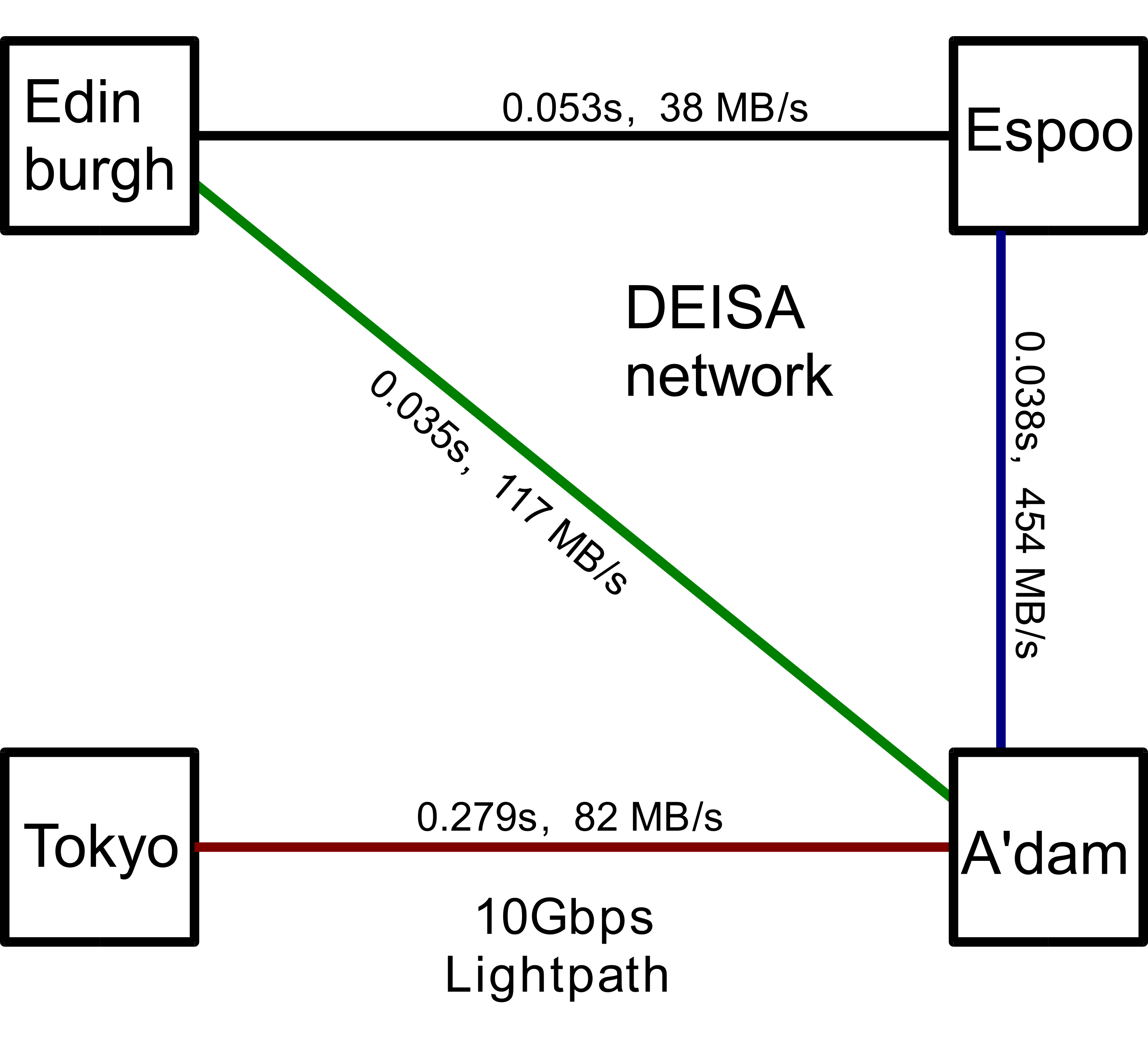} &
  \includegraphics[width=0.47\columnwidth]{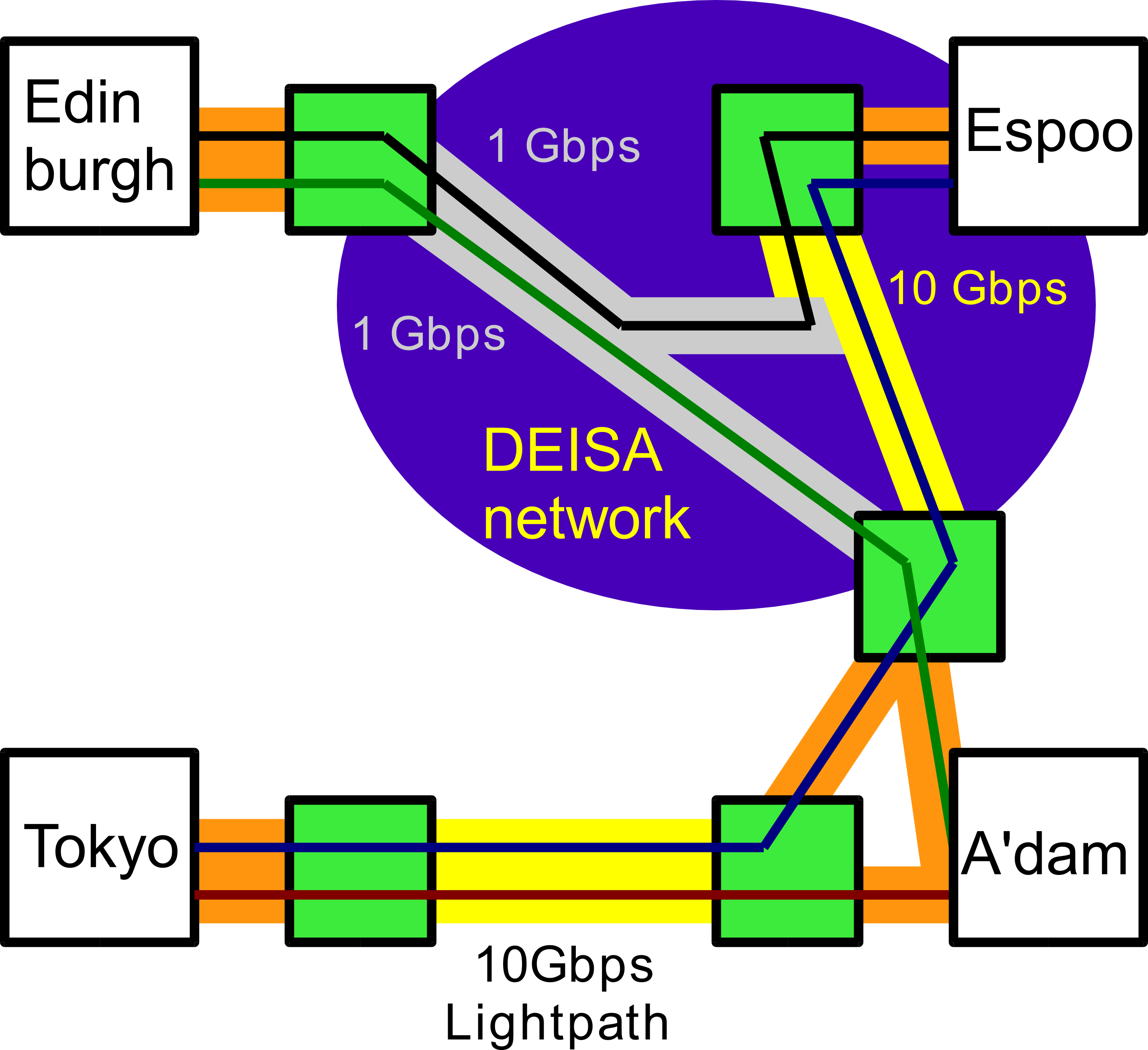} \\
  \end{tabular}
  
  \caption{Network overview of a cosmological simulation across four
supercomputers. The network latency and maximum sustained throughput for 64 MB
message exchanges (using MPWide with the settings described in the text) over
each network path is given on the left, and an overview of the network topology
is given on the right. Here, the communication nodes are indicated by the solid
green boxes.}
  
  \label{Fig:GBBPtopology}
\end{figure}

Since most of the supercomputers are connected to the high speed
network through specialized communication nodes, we are required to forward our
messages through these nodes to exchange data between supercomputers. This
forwarding is done in user space with MPWide Forwarder programs. A
graphical overview of the network topology, including the communication nodes as
well as latency and bandwidth characteristics for each network path, can
be found in Fig.~\ref{Fig:GBBPtopology}.

\begin{table}
\centering
    \begin{tabular}{|l|l|l|l|l|}

\hline
Name                 & Huygens    & Louhi    & HECToR    & CFCA\\
\hline
Location             & A'dam      & Espoo    & Edinburgh & Tokyo \\
Vendor               & IBM        & Cray     & Cray      & Cray  \\
Architecture         & Power6     & XT4      & XT4       & XT4 \\
\# of nodes          & 104        & 1012     & 1416      & 740 \\
Cores per node       & 32         & 4        & 16        & 4 \\
CPU [GHz]            & 4.7        & 2.3      & 2.3       & 2.2 \\
RAM / core [GB]      & 4/8        & 1/2      & 2         & 2 \\
WAN [Gbps]           & 2x10       & 10       & 1         & 10 \\
Peak [TFLOP/s]       & 64.97      & 102.00   & 208.44    & 28.58 \\
Order in plots       & $1^{st}$   & $2^{nd}$ & $3^{rd}$  & $4^{th}$ \\

$\tau_{\rm tree}$ [$\times 10^{-9}$ s] & 5.4 & 3.9 & 4.0 & 4.3 \\
$\tau_{\rm fft}$  [$\times 10^{-9}$ s] & 5.1 & 3.4 & 3.4 & 3.4 \\
$\tau_{\rm mesh}$ [$\times 10^{-7}$ s] & 5.8 & 7.8 & 7.8 & 7.8 \\
\hline

    \end{tabular}

\caption{Technical specifications of the IBM Power 6 supercomputer in
Amsterdam (The Netherlands) and the Cray-XT4 supercomputers in Espoo (Finland),
Tokyo (Japan) and Edinburgh (United Kingdom). Note that Huygens is connected to
both the 10Gbps DEISA network and the 10Gbps light path between Amsterdam and
Tokyo.}

\label{Tab:Superspecs}
\end{table}

\begin{table}
\centering
  \begin{tabular}{llrrr}
  \hline
  Name                  & Description         & DAS-3 Value               & GBBP Value                & unit \\
  \hline
  $\lambda_{\rm lan}$   & LAN round-trip time & $1.0 \times 10^{-4}$      & $8.0 \times 10^{-5}$      & [s] \\
  $\lambda_{\rm wan}$   & WAN round-trip time & $3.0 \times 10^{-3}$      & $2.7 \times 10^{-1}$      & [s] \\
  $\sigma_{\rm lan}$    & LAN bandwidth       & $1.0 \times 10^{8}$       & $5.4 \times 10^{8}$       & [bytes/s]\\
  $\sigma_{\rm wan}$    & WAN bandwidth       & $5.0 \times 10^{7} / s-1$ & $5.0 \times 10^{7}$       & [bytes/s]\\
  \hline
  \end{tabular}

  \caption{List of network parameters used for modelling the performance of our
  runs. The name of the constant is given in the first column, followed by a
  brief description of the constant in the second column, the value used for
  modelling the DAS-3 runs in the third column, the value used for modelling
  the GBBP runs in the fourth column and the unit used in the fifth column.
  $\sigma_{\rm lan}$ for GBBP was based on the point-to-point bandwidth per core
  on HECToR. Since the wide area network in the DAS-3 resembles a star topology, 
  we divide the available bandwidth by $s-1$ there.}

\label{Tab:PerfConstants}

\end{table}

\subsection{GBBP results}

\begin{figure}
\centering
  \includegraphics[width=0.8\columnwidth]{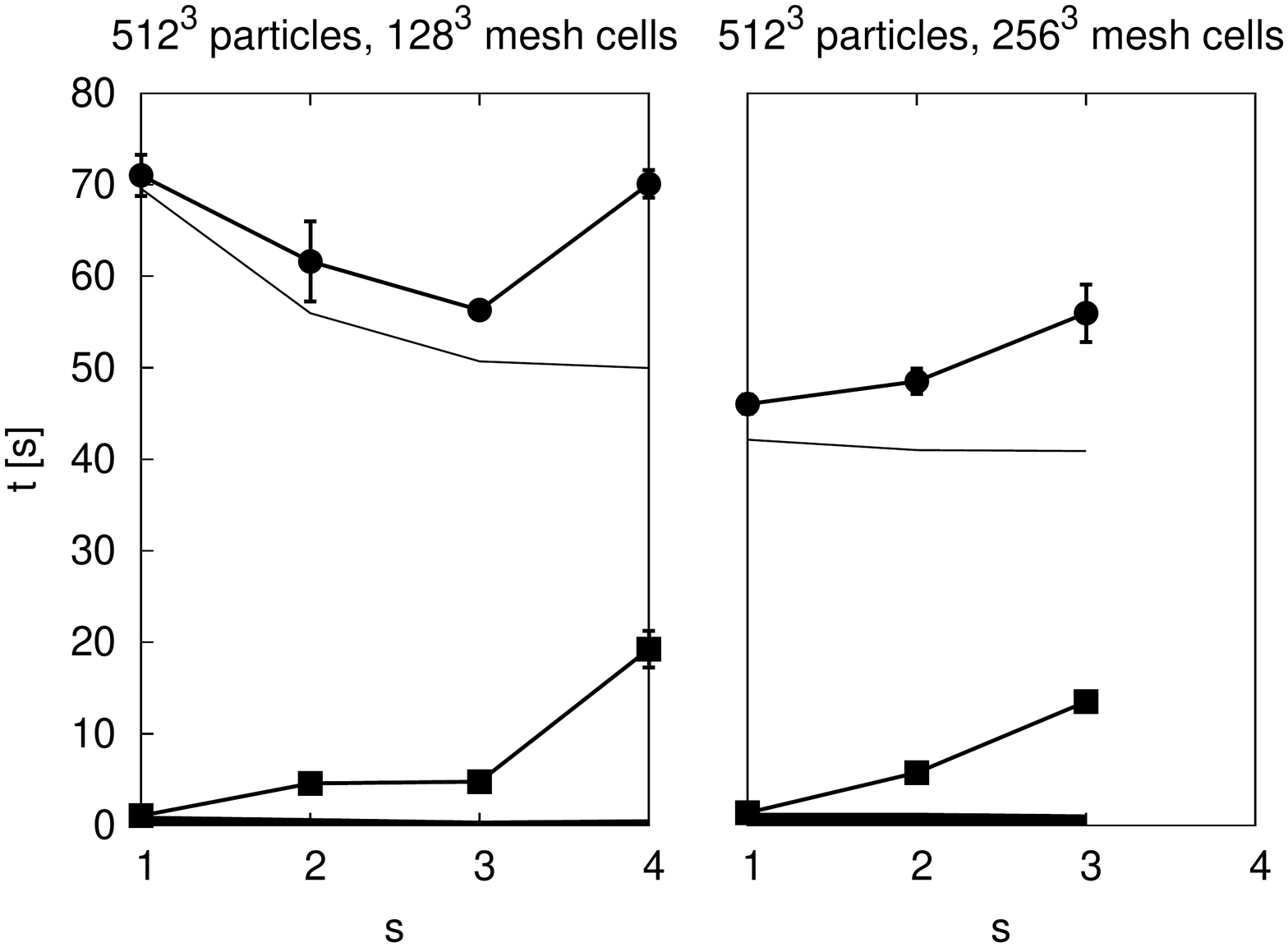}

\caption{As Fig.~\ref{Fig:Das3runs}, but these simulations were run using 120
cores in total on the planet-wide network of supercomputers. The wall-clock
time is given as a function of the number of supercomputers $s$. The average
time spent on PM integration was 0.89-1.24 s for the runs using $M=128^3$, and
3.44-3.78 s for the runs using $M=256^3$. The ordering and technical
specifications of the supercomputers we used are given in
Tab.~\ref{Tab:Superspecs}.}

\label{Fig:GBBPruns}
\end{figure}

\begin{figure}
\centering
  \includegraphics[width=0.8\columnwidth]{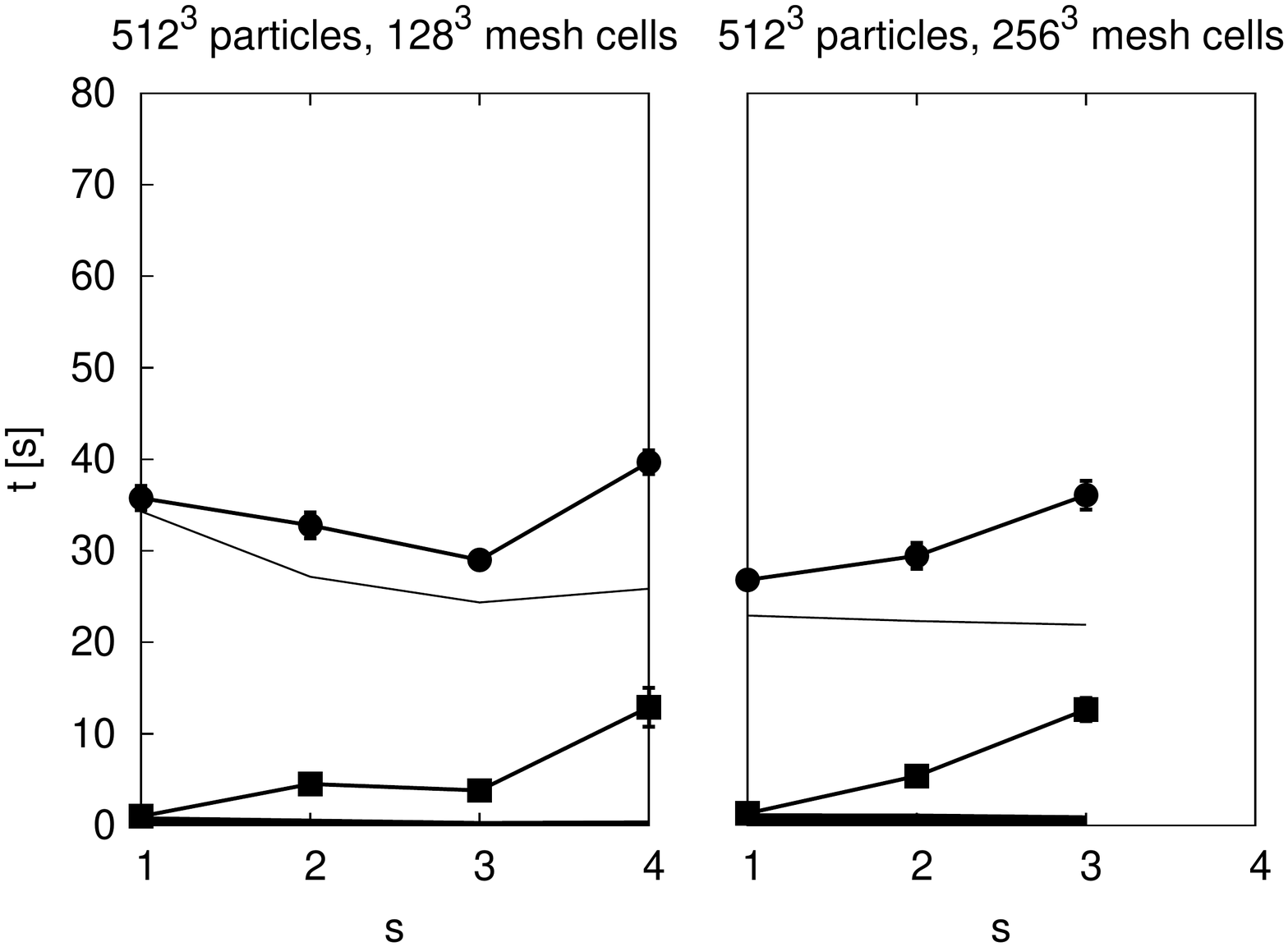}

\caption{As Fig.~\ref{Fig:GBBPruns}, , but for simulations using
      $\theta=0.5$. The average time spent on PM integration was 
      0.89-1.24 s for the runs using $M=128^3$, and
      3.44-3.78 s for the runs using $M=256^3$.}

\label{Fig:GBBPruns2}
\end{figure}

The timing measurements of several simulations using $N=256^3$ are given in the
right panel of Fig.~\ref{Fig:256runs} and \ref{Fig:256runs2} and measurements of simulations using
$N=512^3$ are given in Fig.~\ref{Fig:GBBPruns} and \ref{Fig:GBBPruns2}. We provide the timing results
and model predictions for all our experiments in Tables \ref{Tab:GBBPruns03} and
\ref{Tab:GBBPruns05}.

\begin{figure}
  \begin{center}

  \begin{tabular}{cc}
    \includegraphics[width=0.47\columnwidth]{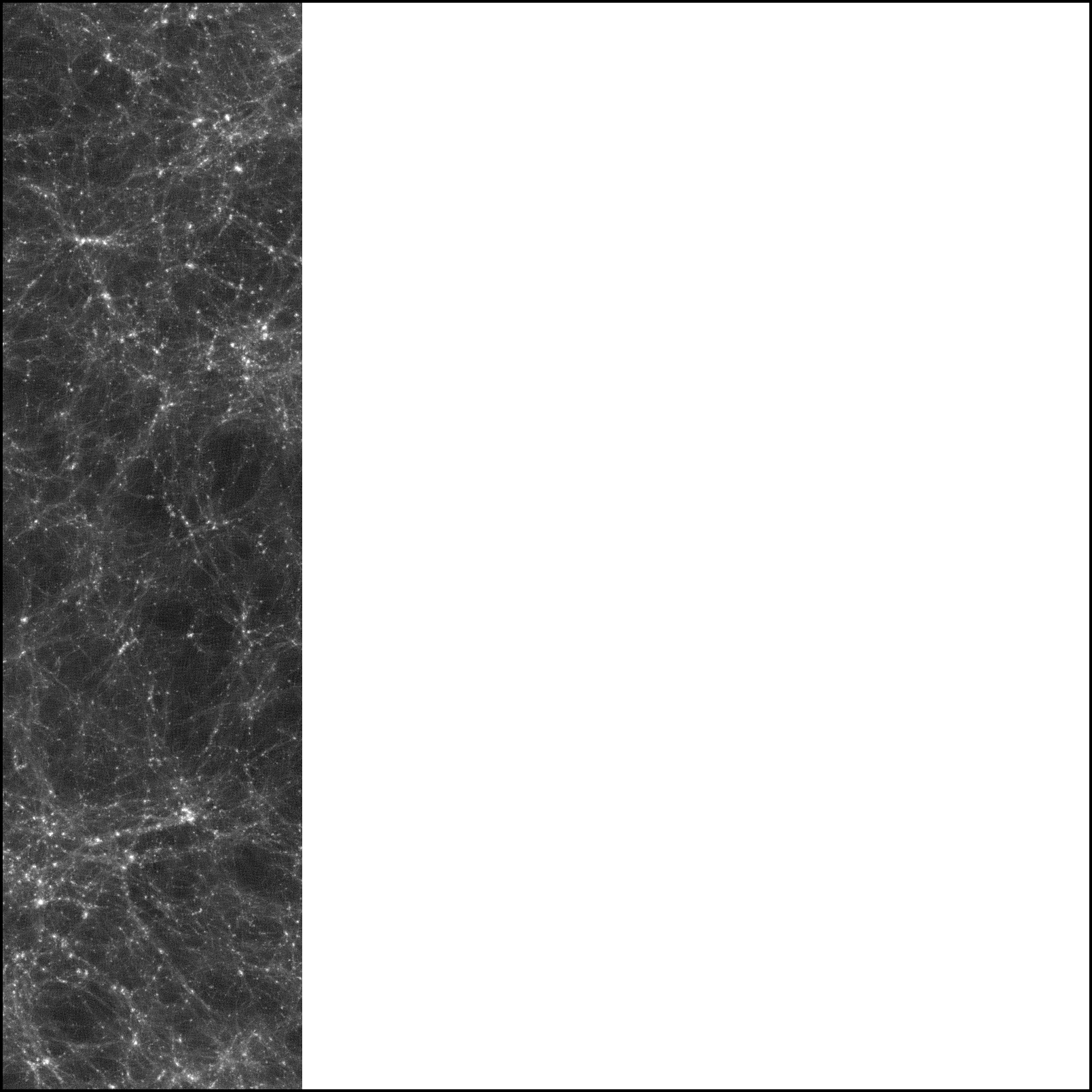} &
    \includegraphics[width=0.47\columnwidth]{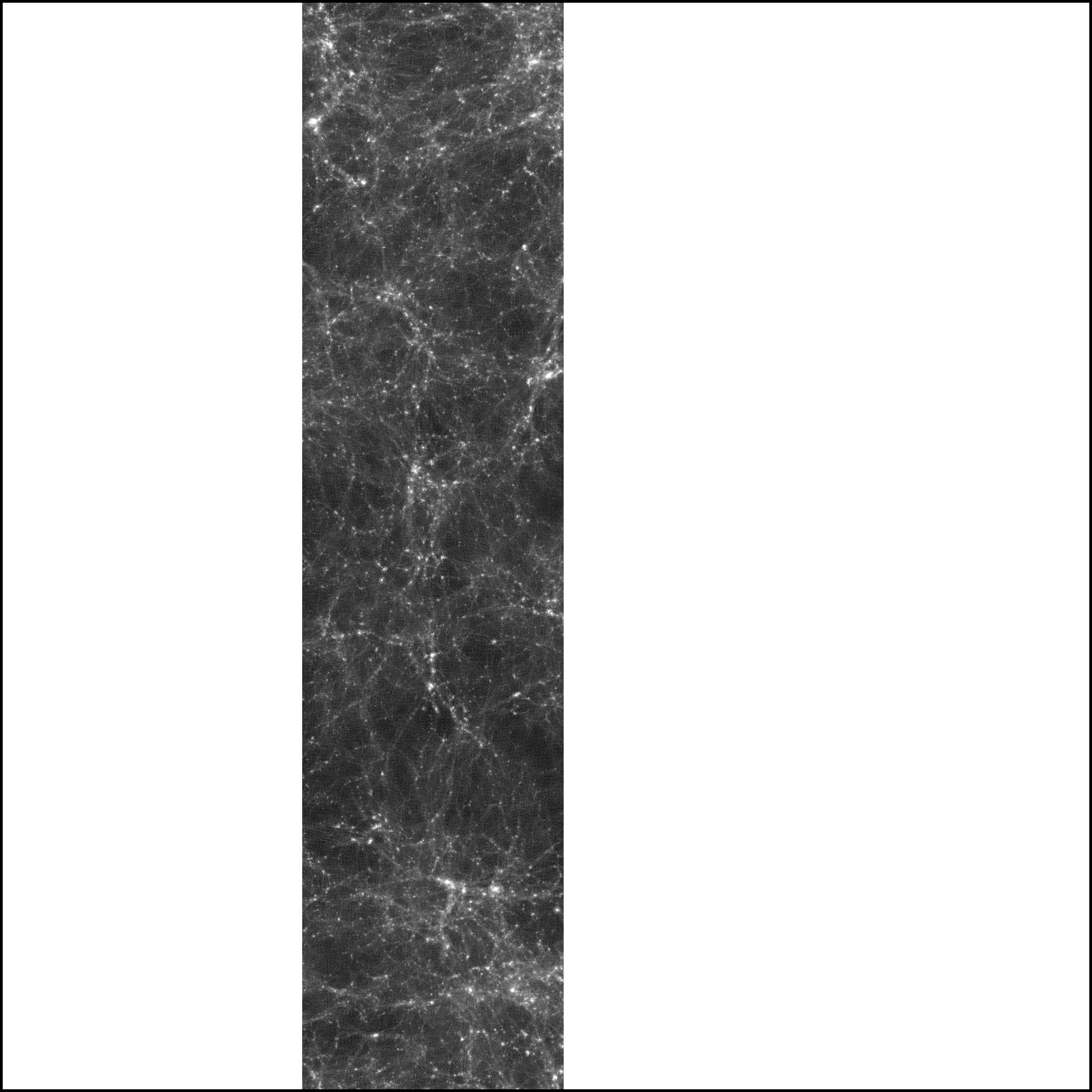} \\
    \includegraphics[width=0.47\columnwidth]{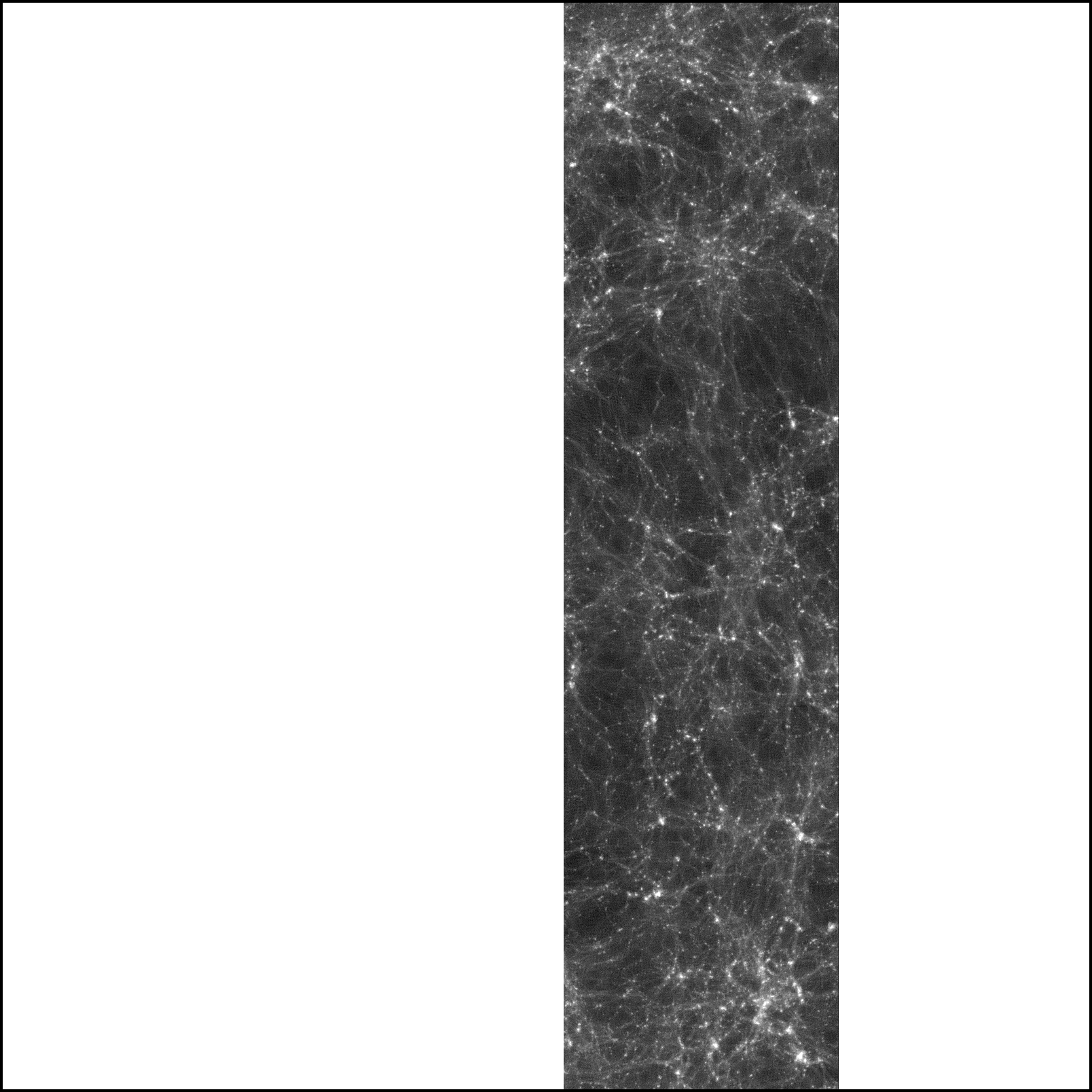} &
    \includegraphics[width=0.47\columnwidth]{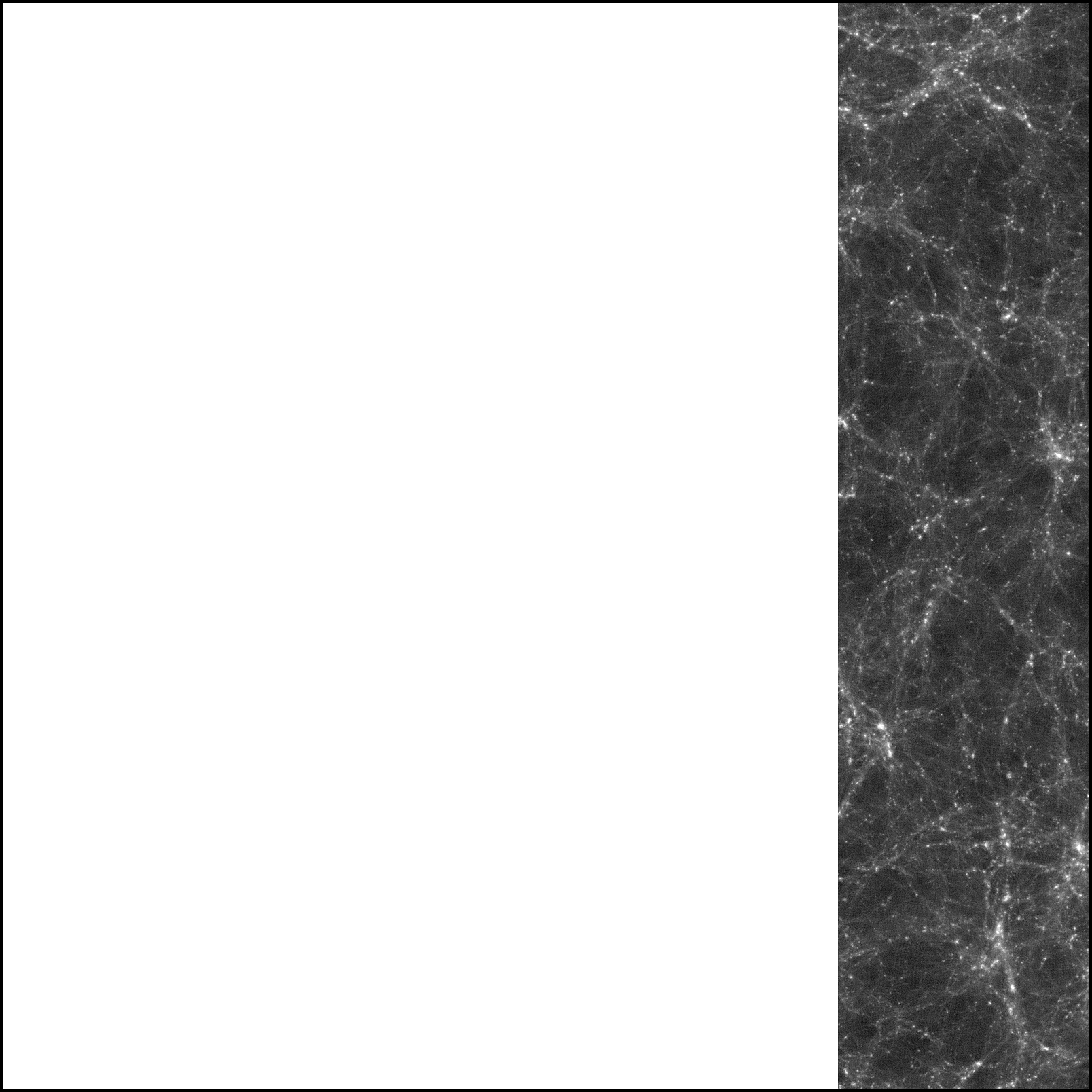} \\
  \end{tabular}

\caption{Snapshots at $z=5.65$ of a simulation with $N=512^3$ over four
supercomputers. The slices have been placed to match the volumes on the
supercomputers that reside respectively (from top left to bottom right) 
in Tokyo, Espoo, Edinburgh and Amsterdam.}

\label{Fig:GBBPsnapshots}
  \end{center}
\end{figure}

We have performed a number of runs with $N=256^3$ across up to three
supercomputers. The runs over two supercomputers in the DEISA network have 
a communication overhead between 1.08 and 1.54 seconds. This constitutes between 8 and 13\% of the
total runtime for runs with $\theta=0.3$ and between 15 and 20\% for runs with
$\theta=0.5$. The run with $N=256^3$ between Edinburgh and Amsterdam has 
less communication overhead than the run between Espoo and Amsterdam, despite
the 1 Gbps bandwidth limitation on the connection to Edinburgh. Our run
between Amsterdam and Espoo suffered from a high background load,
which caused the communication time to fluctuate over 10 steps with $\sigma
\sim 0.25$ s, compared to $\sigma \sim 0.04$ s for the run between Amsterdam
and Edinburgh. The run over three sites has a higher overhead than the runs 
over two sites, mainly due to using the direct connection between Edinburgh 
and Espoo, which is poorly optimized.

The wall-clock time of the simulations with $N=512^3$ is generally dominated by
calculations, although the communication overhead becomes higher as we increase
$s$. The runs over two sites (Espoo and Amsterdam) spend about
4-5 seconds on communication, which is less than 10 percent of the total
execution time for the run using $\theta=0.3$. For simulations with $M=128^3$,
the use of three supercomputers rather than two does not significantly increase
the communication overhead. However, when we run simulations with $N=512^3$ and
$M=256^3$ the use of a third supercomputer doubles the communication overhead.
This increase in overhead can be attributed to the larger mesh size, as the now
larger data volume of the mesh exchange scales with $s$.

\begin{figure}
\begin{center}
  \includegraphics[width=\columnwidth]{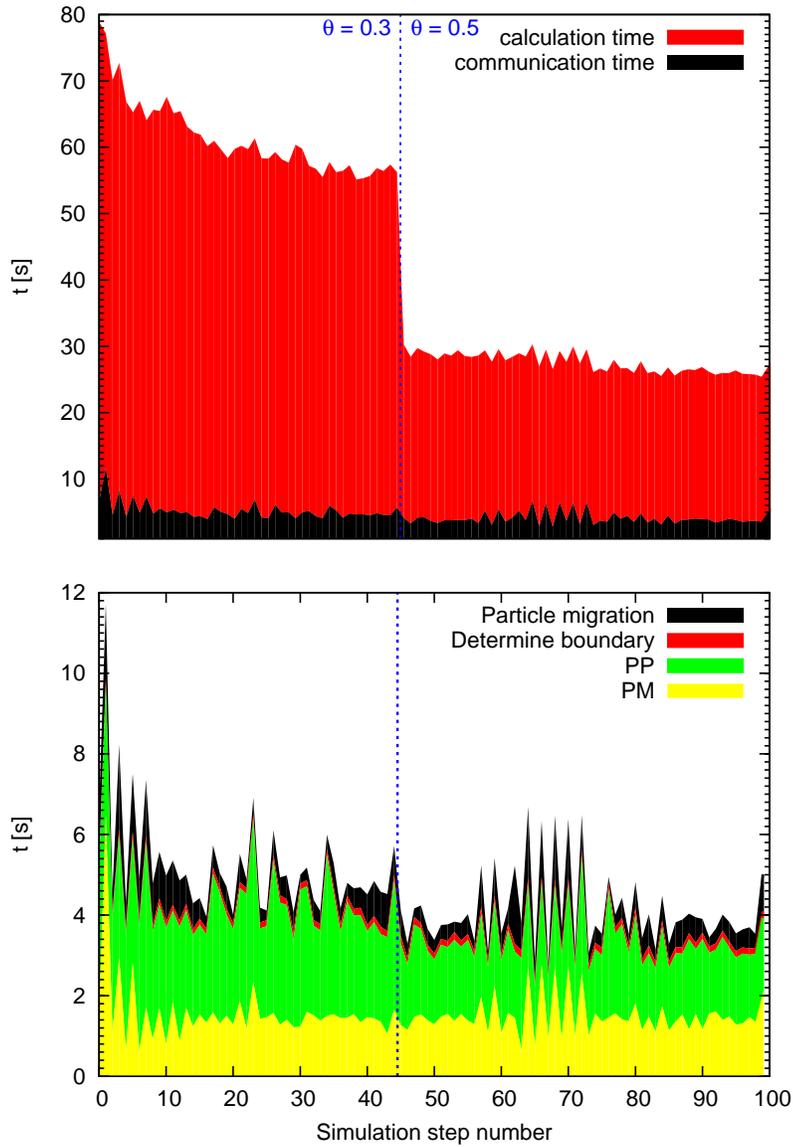}

  \caption{Performance results of a simulation with $N=512^3$ and $M=128^3$
  using 120 processes in total across supercomputers in Amsterdam, Edinburgh
  and Espoo as a function of the simulation step number. The wall-clock time
  spent per step on calculations (top red area) and communications (bottom
  black area) is given in the top figure. Time spent on the four communication
  phases is given in the bottom figure. These phases are (from top to bottom)
  the migration of particles between sites, the exchanges of sample particles 
  for determining the site boundaries, the local essential tree exchanges (PP)
  and the mesh cell exchanges (PM).}

    \label{Fig:GBBPexamplerun}
\end{center}
\end{figure}

We provide an overview of the time spent per simulation step for the run with
$N=512^3$ and $M=128^3$ over three sites (Edinburgh, Espoo and
Amsterdam) in Fig.~\ref{Fig:GBBPexamplerun}. For this run the total 
communication overhead of the simulation code remains limited and relatively 
stable, achieving slightly lower values at later steps where $\theta=0.5$. 
A decomposition of the communication overhead for this run is given in
the bottom panel of Fig.~\ref{Fig:GBBPexamplerun}. The time required for exchanging the local
essential tree constitutes about half of the total communication overhead.
The simulation changes the opening angle at step 46, from $\theta=0.3$ to
$\theta=0.5$. As a result, the size of the local essential tree becomes 
smaller and less time is spent on exchanging the local essential trees.
The time spent on the other communications remains roughly constant throughout 
this run. 

We have had a brief opportunity to run across all four supercomputers, which
allowed us to run a simulation with $N=512^3$. We provide a density plot
of the particle data present at each site at the end of this run in
Fig.~\ref{Fig:GBBPsnapshots}. In addition, the timing measurements of the run 
across four sites are given in the left panel of Fig.~\ref{Fig:GBBPruns}.
Here we observe an increase in communication time from 3.8-4.8 seconds for
a run over the three DEISA sites to 12.9-19.3 seconds for the run over all four
sites. However, the runs over three sites relay messages between
Edinburgh and Espoo through the supercomputer in Amsterdam, whereas the four
site run uses the direct connection between these sites, which is poorly
optimized. If we use the direct connection for a run across three sites the
communication overhead increases from 3.8-4.8 seconds to 10.1-13.2 seconds per
step (see Tables \ref{Tab:GBBPruns03} and \ref{Tab:GBBPruns05}). Therefore, 
the communication overhead of a run across four sites may be reduced by 
a factor of two if we would relay the messages between Espoo and Edinburgh over 
Amsterdam. 


\begin{table}
\centering 
    \begin{tabular}{llllllllllll}

\hline
$N^{1/3}$ & $M^{1/3}$ & np & $s$ & \multicolumn{2}{c}{comm.} & tree & exec. & \multicolumn{2}{c}{$t_{\rm comm}$} & $t_{\rm tree}$ & $t_{\rm exec}$ \\

     &     &     &        & local  & total &      &        & only  & $+w_{comm}$ &      &      \\
\hline
     &     &     &        & real   & real  & real & real   & model & model       & model & model\\
     &     &     & \#     & [s]    & [s]   & [s] & [s]     & [s]   & [s]         & [s]    & [s] \\
\hline
256  & 128 & 60  & A     & 0.207 & 0.207 & 11.32 & 11.80   & 0.04   & 0.03  & 10.79 & 11.22 \\
256  & 128 & 60  & HA    & 0.156 & 1.085 & 10.88 & 12.23   & 0.03   & 1.06  & 9.29 & 10.74 \\
256  & 128 & 60  & EA    & 0.177 & 1.536 & 10.91 & 12.71   & 0.03   & 1.06  & 9.39 & 10.84 \\ 
256  & 128 & 60  & AT    & 0.177 & 3.741 & 11.47 & 15.48   & 0.03   & 4.23  & 9.69 & 14.31 \\
256  & 128 & 60  & HEA*  & 0.143 & 5.960 & 10.95 & 17.12   & 0.03   & 1.66  & 8.86 & 10.91 \\
\hline
512  & 128 & 120 & A     & 1.045 & 1.045 & 69.57 & 71.04   & 0.06   & 0.06  & 58.98 & 59.91 \\
512  & 128 & 120 & HA    & 0.783 & 4.585 & 55.96 & 61.62   & 0.04   & 3.14  & 50.79 & 54.91 \\
512  & 128 & 120 & HEA   & 0.524 & 4.778 & 50.69 & 56.31   & 0.04   & 4.19  & 48.42 & 53.64 \\
512  & 128 & 120 & HEA*  & 0.760 & 13.17 & 56.82 & 71.01   & 0.04   & 4.19  & 48.42 & 53.64 \\ 
512  & 128 & 120 & HEAT* & 0.653 & 19.26 & 49.96 & 70.10   & 0.04   & 9.42  & 48.06 & 58.52 \\
\hline
512  & 256 & 120 & A     & 1.367 & 1.367 & 42.13 & 46.04   & 0.16   & 0.16  & 49.60 & 52.46 \\
512  & 256 & 120 & HA    & 1.363 & 5.761 & 41.00 & 48.52   & 0.15   & 5.48  & 42.71 & 51.01 \\
512  & 256 & 120 & EA    & 1.247 & 7.091 & 37.60 & 46.37   & 0.15   & 5.48  & 43.17 & 51.47 \\
512  & 256 & 120 & AT    & 1.313 & 10.98 & 38.97 & 51.67   & 0.15   & 8.66  & 44.54 & 56.02 \\
512  & 256 & 120 & HEA   & 1.191 & 13.51 & 40.91 & 55.96   & 0.14   & 7.66  & 40.72 & 51.22 \\
\hline
1024 & 256 & 240 & E     & 3.215 & 3.215 & 182.4 & 189.6   & 0.20 & 0.20  & 200.7 & 205.8 \\
1024 & 256 & 240 & A     & 3.402 & 3.402 & 265.5 & 272.8   & 0.20 & 0.20  & 271.0 & 275.8 \\
1024 & 256 & 240 & HA    & 3.919 & 21.98 & 217.9 & 252.0   & 0.18 & 23.88 & 233.4 & 262.3 \\
1024 & 256 & 240 & HEA   & 4.052 & 31.28 & 258.0 & 294.7   & 0.17 & 26.05 & 217.1 & 248.4 \\
\hline
      \end{tabular}

  \caption{List of runs performed for GBBP that use $\theta=0.3$. The cube root
of the number of particles and mesh cells are given in the first and second
column, the number of processes and the supercomputers involved in the third
and fourth column. Here, the letters correspond to supercomputers in Espoo
(given by ``H"), (E)dinburgh, (A)msterdam and (T)okyo. For runs over three and
four sites, we relayed all traffic through the supercomputer in Amsterdam
except for the runs marked with an asterisk, which used the direct network
between Edinburgh and Espoo. The next four columns contain timing measurements
from our experiments, which are average times per step averaged over 10 steps.
The columns contain respectively the intra-site communication time, the total
communication time, the time spent on tree integration (excluding PM) and the
total wall-clock time. The last four columns contain the intra-site
communication time, total communication time, tree integration time (excluding PM), and
wall-clock time as predicted by our performance model.}

\label{Tab:GBBPruns03}
\end{table}

\begin{table}
\centering
    \begin{tabular}{llllllllllll}

\hline
$N^{1/3}$ & $M^{1/3}$ & np & $s$ & \multicolumn{2}{c}{comm.} & tree & exec. & \multicolumn{2}{c}{$t_{\rm comm}$} & $t_{\rm tree}$ & $t_{\rm exec}$ \\

     &     &     &        & local  & total &      &        & only  & $+w_{comm}$ &      &      \\
\hline
     &     &     &        & real   & real  & real & real   & model & model       & model & model\\
     &     &     & \#     & [s]    & [s]   & [s] & [s]     & [s]   & [s]         & [s]    & [s] \\
\hline
256  & 128 & 60  & A       & 0.197 & 0.197 & 5.968 & 6.446   & 0.04 & 0.04 & 5.42 & 5.84 \\
256  & 128 & 60  & HA      & 0.156 & 1.043 & 5.786 & 7.087   & 0.03 & 0.98 & 4.66 & 6.03 \\
256  & 128 & 60  & EA      & 0.159 & 1.480 & 5.827 & 7.567   & 0.03 & 0.98 & 4.71 & 6.08 \\
256  & 128 & 60  & AT      & 0.157 & 3.575 & 6.024 & 9.858   & 0.03 & 4.15 & 4.86 & 9.40 \\
256  & 128 & 60  & HEA*    & 0.143 & 5.827 & 6.047 & 12.08   & 0.03 & 1.58 & 4.45 & 6.41 \\
\hline
512  & 128 & 120 & A       & 0.992 & 0.992 & 34.31 & 35.76   & 0.05 & 0.05 & 29.59 & 30.52 \\
512  & 128 & 120 & HA      & 0.732 & 4.522 & 27.16 & 32.78   & 0.04 & 2.82 & 25.48 & 29.29 \\
512  & 128 & 120 & HEA     & 0.467 & 3.803 & 24.35 & 28.99   & 0.04 & 3.87 & 24.30 & 29.19 \\
512  & 128 & 120 & HEA*    & 0.760 & 10.86 & 28.52 & 40.45   & 0.04 & 3.87 & 24.30 & 29.19 \\ 
512  & 128 & 120 & HEAT*   & 0.534 & 12.90 & 25.84 & 39.68   & 0.03 & 9.09 & 24.11 & 34.25 \\
\hline
512  & 256 & 120 & A       & 1.308 & 1.308 & 22.91 & 26.80   & 0.16 & 0.16 & 24.89 & 27.74 \\
512  & 256 & 120 & HA      & 1.284 & 5.383 & 22.32 & 29.45   & 0.14 & 5.16 & 21.43 & 29.40 \\
512  & 256 & 120 & EA      & 1.213 & 6.991 & 20.75 & 29.44   & 0.14 & 5.16 & 21.66 & 29.63 \\
512  & 256 & 120 & AT      & 1.223 & 9.640 & 20.88 & 32.22   & 0.14 & 8.33 & 22.35 & 33.50 \\
512  & 256 & 120 & HEA     & 1.100 & 12.65 & 21.92 & 36.08   & 0.14 & 7.33 & 20.43 & 30.61 \\
\hline
1024 & 256 & 240 & HA      & 3.490 & 19.06 & 104.1 & 128.2   & 0.17 & 22.59 & 117.1 & 144.8 \\
\hline
2048 & 256 & 750 & AT      & 16.58 & 46.50 & 413.9 & 483.4   & 0.26 & 17.59 & 443.2 & 470.3 \\
\hline
    \end{tabular}

\caption{As Tab.~\ref{Tab:GBBPruns03} but with $\theta=0.5$ instead
of $\theta=0.3$. The run with $N=2048^3$ used an older
version of the code and different settings, and is described in 
\cite{CosmoGrid}.}

  \label{Tab:GBBPruns05}
\end{table}

We also ran several simulations with $N=1024^3$. The load balancing
increases the communication volume of these larger runs considerably, as 850MB
of particle data was exchanged on average during each step. However, the
communication time still constitutes only one-tenth of the total run-time for
the simulation across three supercomputers. 

\section{Scalability of $N$-body simulations across supercomputers}\label{GreeMgridmodel}

In this section we use our performance model and simulation results to predict
the scalability of $N$-body code across grid sites. In the first part of this
section we examine the predicted speedup and efficiency when scaling three
cosmological $N$-body problems across multiple supercomputers. In the second
part we apply existing performance models for tree codes (with block and shared
time step schemes) and direct-method codes to predict the scalability of three
stellar $N$-body problems across supercomputers. In the third part we predict
the efficiency of cosmological $N$-body simulations over 8 sites as a function
of the available bandwidth between supercomputers. We provide an overview of
the three cosmological problems in Tab.~\ref{Tab:Nprobs}, and an overview of
the three stellar problems in Tab.~\ref{Tab:Nprobs2}.

\begin{table}[b]
\centering
\begin{tabular}{ l l l l l}
\hline
Integrator & $N$      & $M$      & np 1 & np 2 \\
\hline
TreePM     & $2048^3$ & $256^3$  & 128  & 2048  \\
TreePM     & $2048^3$ & $1024^3$ & 128  & 2048  \\
TreePM     & $8192^3$ & $1024^3$ & 4096 & 32768 \\
\hline
\end{tabular}

\caption{Description of the cosmological $N$-body problems used for the scalability
analysis across supercomputers. The type of integrator is given in
the first column, the number of particles in the second column and the number of
mesh cells in the third column. The last two columns contain respectively the
number of processes per site for the speedup analysis and the number of total
processes for the efficiency and bandwidth analysis.}

\label{Tab:Nprobs}
\end{table}

These problems are mapped to a global grid infrastructure, which has 
a network latency of 0.3 s and a bandwidth capacity of 400 MB/s
between supercomputers. The machine constants are similar to the ones
we used for our runs, and can be found in Tab.~\ref{Tab:GridConstants}. 
To limit the complexity of our analysis, we 
assume an identical calculation speed for all cores on all sites. Our
performance predictions use an opening angle $\theta=0.5$.

\begin{table}
\centering
  \begin{tabular}{ l l l r }
  \hline
  Name of constant     & Value                & unit \\
  \hline
  $\tau_{\rm tree}$    & $5.0 \times 10^{-9}$ & [s]\\
  $\tau_{\rm fft}$     & $3.5 \times 10^{-9}$ & [s]\\
  $\tau_{\rm mesh}$    & $7.5 \times 10^{-7}$ & [s]\\
  $\lambda_{\rm lan}$  & $8.0 \times 10^{-5}$ & [s] \\
  $\lambda_{\rm wan}$  & $3.0 \times 10^{-1}$ & [s] \\
  $\sigma_{\rm lan}$   & $2.3 \times 10^{9}$  & [bytes/s]\\
  $\sigma_{\rm wan}$   & $4.0 \times 10^{8}$  & [bytes/s]\\
  \hline
  \end{tabular}
\caption{List of network parameters used for the scalability
  predictions of our code. The name
  of the constant can be found in the first column, the
  value used for our global grid model in the second 
  column and the units used for each value in the third column.}

  \label{Tab:GridConstants}
\end{table}

\subsection{Speedup and efficiency predictions for TreePM simulations}

\begin{figure}
  \includegraphics[width=0.9\columnwidth]{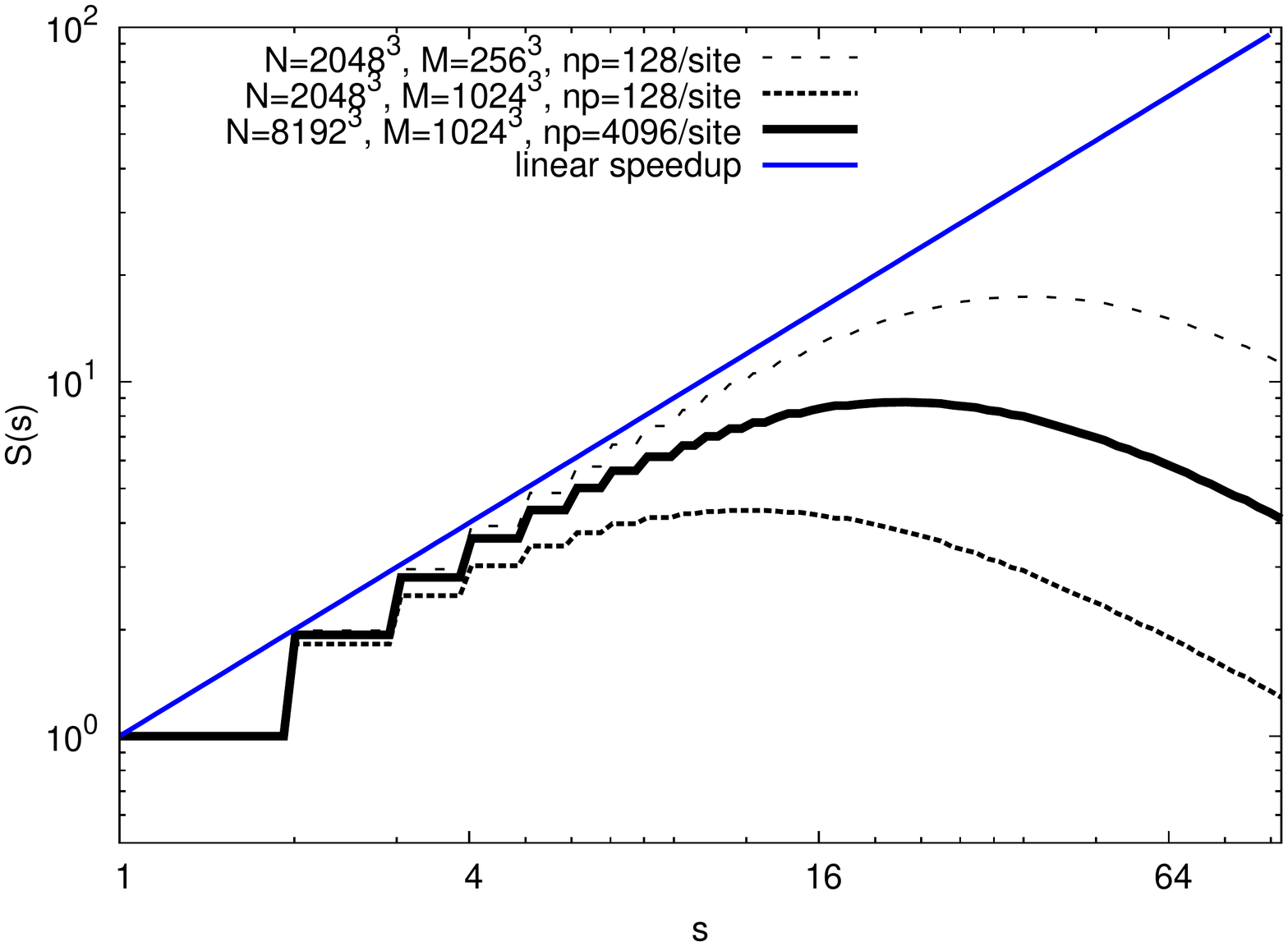}

  \caption{Predicted speedup $S(s)$ of simulations using the TreePM method as a function of the
  number of sites $s$ in a global grid. The total number of processes scales linearly
  with $s$.}

  \label{Fig:DiscPerfG}
\end{figure}

The predicted speedup $S(s)$ (as defined in Eq.~\ref{Eq:sp}) for three
cosmological simulations as a function of the number of supercomputers $s$ can
be found in Fig.~\ref{Fig:DiscPerfG}. As the number of processes per site
remains fixed, the total number of processes $p$ increases linearly with $s$.
All example problems run efficiently over up to 3 sites. The simulation with
$N=2048^3$ and $M=256^3$ scales well as $s$ increases and obtains a speedup of
$\sim 13$ when $s=16$. When scaling up beyond $s \sim 25$, the speedup
diminishes as the simulation becomes dominated by communication. 

The simulation with $N=2048^3$ and $M=1024^3$ does not scale as well and only
achieves a good speedup when run across a few sites. Here, the speedup as
a function of $s$ begins to flatten at $s \sim 5$, due to the
serial integration of the larger mesh. For $s \apgt 16$, the communication
overhead begins to dominate performance and the speedup decreases for higher $s$. 
The speedup of the run with $N=8192^3$ and $M=1024^3$ scales better
with $s$ than the speedup of the run with $N=2048^3$ because it spends more
time per step on tree force calculations.

\begin{figure}
  \includegraphics[width=0.9\columnwidth]{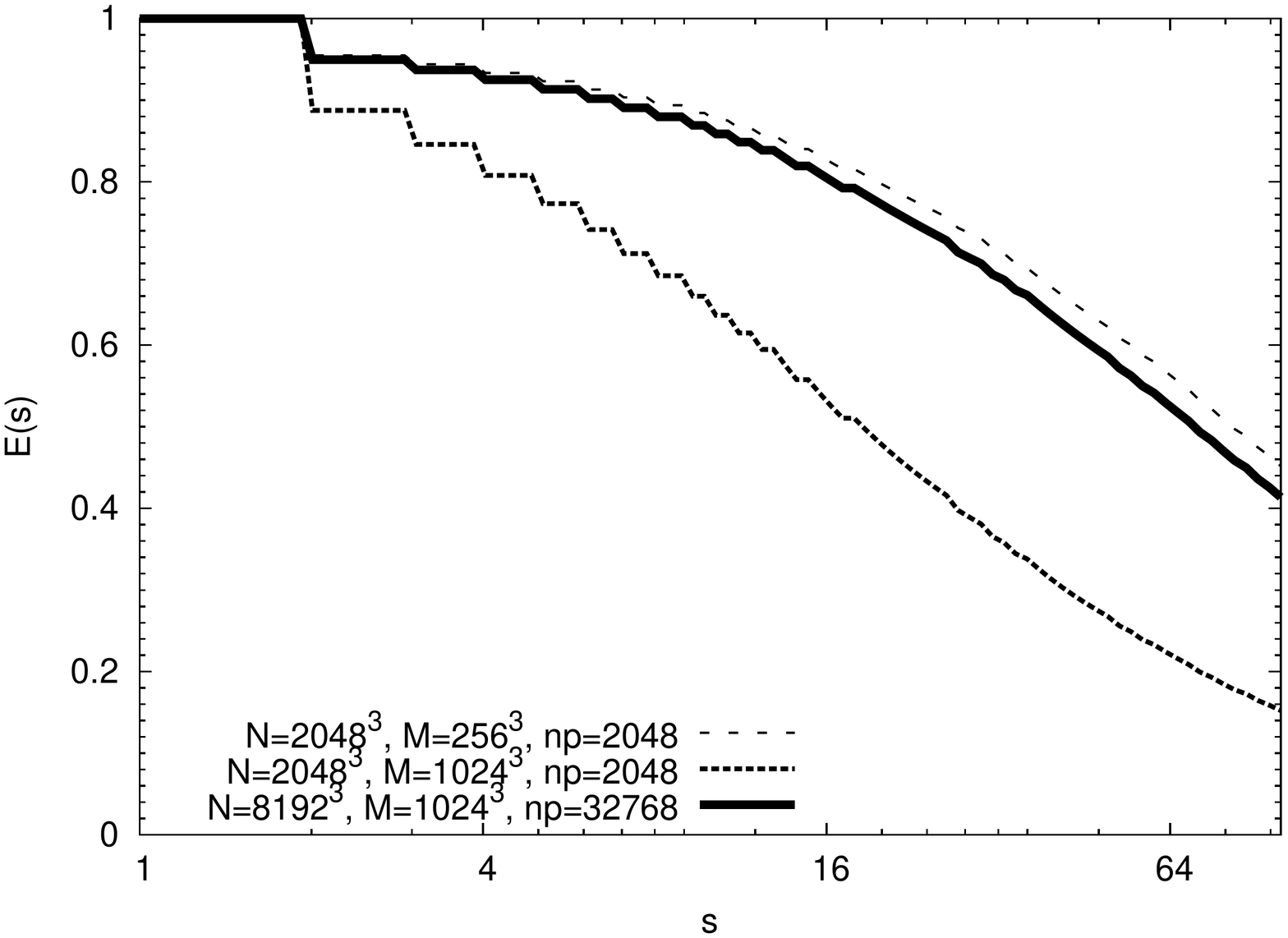}
  
  \caption{Predicted efficiency $E(s)$ of simulations using the TreePM method as a function of the
  number of sites $s$ in a global grid. The total number of processes is
  kept fixed for all simulations.}
  
  \label{Fig:DiscPerfGeff}
\end{figure}

We provide the predicted efficiency of three simulations over $s$
supercomputers relative to a simulation over one supercomputer, $E(s)$, using
the same number of processes in Fig.~\ref{Fig:DiscPerfGeff}. Here, the run with
$N=2048^3$ and $M=256^3$ and the run with $N=8192^3$ and $M=1024^3$ retain a
similar efficiency as we scale up with $s$. If $s \apgt
6$, the run with $N=8192^3$ is slightly less efficient than the
simulation with $N=2048^3$ and $M=256^3$. The simulation using $N=2048^3$ and
$M=1024^3$ is less efficient than the other two runs. The data volume of the
mesh cell exchange is 64 times higher than that of the run with
$M=256^3$, which results in an increased communication overhead.



\subsection{Speedup and efficiency predictions for tree and direct-method simulations}

We present predictions of three example $N$-body simulations of stellar
systems, each of which uses a different integration method. The integration
methods we use in our models are a Barnes-Hut tree algorithm
\cite{1986Natur.324..446B} with shared time steps, a tree algorithm using block
time steps \cite{1986LNP...267.....H} and a direct-method algorithm
\cite{1992PASJ...44..141M} using block time steps. For the tree algorithm we
choose the problem sizes and process counts previously used for the
cosmological simulation models with $N=2048^3$. Modelling the performance of
direct-method simulations using $N=2048^3$ is unrealistic, because one step of
force calculations for the $2048^3$ particles would already take $sim$ 280
years when using 1000 nodes on a PC cluster \cite{Gualandris2007}. We instead
predict the performance of direct-method simulations using a more realistic
problem size of 5 million particles.

We model the tree algorithm using a slightly extended version of the models
presented in \cite{2004PASJ...56..521M,1991PASJ...43..621M} and the
direct-method algorithm using the grid-enabled model presented in
\cite{2008NewA...13..348G}. An overview of the three problems used for our
predictions is given in Tab.~\ref{Tab:Nprobs2}.

\begin{table}
\centering
\begin{tabular}{ l l l l l}
\hline
Integrator & $N$   & np 1 & np 2 & time step scheme\\
\hline
Tree     & $2048^3$& 128 & 2048 & shared\\
Tree     & $2048^3$& 128 & 2048 & block\\
Direct   & $5 \cdot 10^6$ & 16  & 128  & block\\
\hline
\end{tabular}

\caption{Description of the stellar $N$-body problems used for the scalability
analysis across supercomputers. The type of integrator is given in
the first column and the number of particles in the second column. The number of 
processes per site for respectively the speedup analysis and the efficiency analysis 
are given in the third and fourth column. The time step scheme used is given in the 
last column. Note that the direct-method simulation is modelled for
a grid of GPUs.}

\label{Tab:Nprobs2}
\end{table}

The simulations using the tree algorithm are mapped to the same global grid 
infrastructure that we used for modelling the cosmological simulation (see 
Tab.~\ref{Tab:GridConstants} for the machine constants used). The direct-method
simulation is modelled using the network constants from the global grid model,
but here we assume that each node has a Graphics Processing Unit (GPU) with a 
force calculation performance of 200 GFLOP/s. As each force interaction 
requires 66 FLOPs on a GPU, we spend $3.3 \times 10^{-10}$s per force interaction 
(see \cite{2007NewA...12..641P,2009NewA...14..630G} for details on direct 
$N$-body integration on the GPU).

\subsubsection{Performance model for the tree algorithm}

We extend the tree code models given in
\cite{2004PASJ...56..521M,1991PASJ...43..621M} to include the grid
communication overhead by adding the equations for wide area communication of
the tree method from our SUSHI performance model. In accordance with our
previous model, we define the wide area latency-bound time for the tree method
($w_{\rm l,tree}$) as,

\begin{eqnarray}
  w_{\rm l,tree}  &=& \lambda_{\rm wan} \left(4 \left(s-1\right) + 4\right).
\end{eqnarray}

The bandwidth-bound communication time ($w_{\rm b,tree}$) consists of the communication
volume for the local essential tree exchange, which we estimate to be double the size used
in TreePM simulations due to the lack of PM integration, and the communication volume 
required for particle sampling. It is therefore given by,
\begin{eqnarray}
  w_{\rm b,tree} &=& \frac{\left(96/\theta + 48\right) N^{2/3} + 4 N r_{\rm samp}}{\sigma_{\rm wan}},
\end{eqnarray}

where $r_{\rm samp} = 1/10000$ and $\theta=0.5$. We use the equation given in
\cite{1991PASJ...43..621M} for Plummer sphere data sets to calculate the total
number of force interactions. 

\subsubsection{Modelling of block time steps}

Tree and direct $N$-body integrators frequently use a block time step scheme
instead of a shared time step scheme. Block-time step schemes can also be
applied to parallel simulations (unlike e.g., individual time step schemes),
and reduce the computational load by integrating only a subset of particles
during each step \cite{1986LNP...267..233H,1991ApJ...369..200M}. To equalize
the number of integrated particles between simulations, we therefore compare a
single shared time-step integration with $N/n_{\tt b}$ steps of a block
time-step integration, where $n_{\tt b}$ is the average block size. We adopt an
average block size of $n_{\tt b} = 0.2N^{0.81}$ for all block time-step
integrators, the same value that was used in \cite{2008NewA...13..348G}.

\begin{figure}
  \includegraphics[width=0.9\columnwidth]{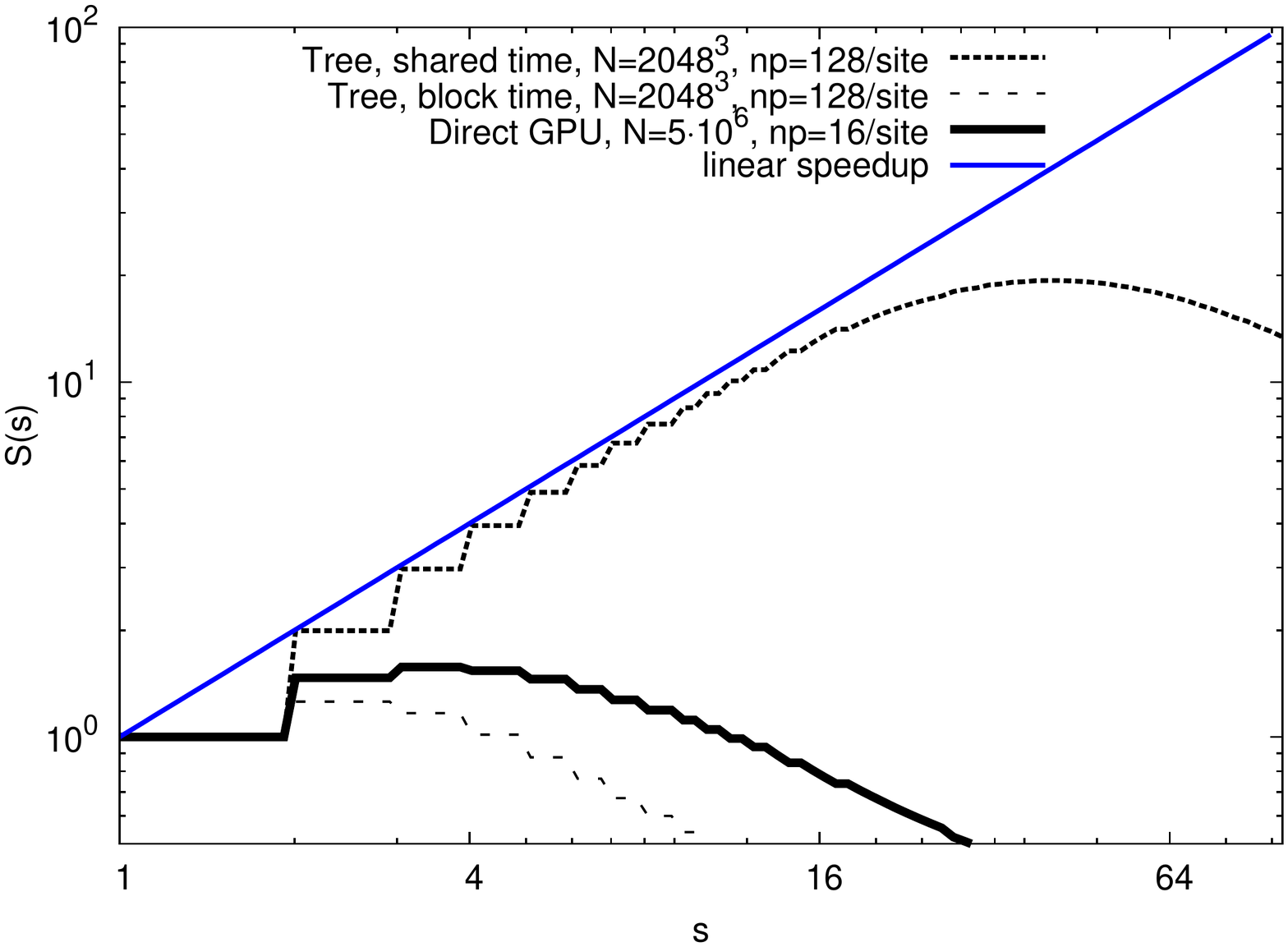}

  \caption{Predicted speedup $S(s)$ of $N$-body simulations using the tree and
  direct method as a function of the number of sites $s$ in a global grid. The
  total number of processes scales linearly with $s$.}

  \label{Fig:Gtree}
\end{figure}

\subsubsection{Predictions}

We give the predicted speedup $S(s)$ of the three simulations as a function of
the number of supercomputers $s$ in Fig.~\ref{Fig:Gtree}. Here, the tree code
with shared time steps scales similarly to the cosmological simulation with
$N=2048^3$ and $M=256^3$ (see Fig.~\ref{Fig:DiscPerfG}). We predict a speedup
of $\sim 13$ when $s=16$. When we model the tree code with a block time step
scheme, the scalability becomes worse because it requires $N/n_{\tt b}$ times
as many communication steps to integrate $N$ particles. The large number of
communications combined with the high latency of wide area networks result in a
high communication overhead. The direct $N$-body run on a global grid of GPUs
also does not scale well over $s$ due to the use of block time steps.

\begin{figure}
  \includegraphics[width=0.9\columnwidth]{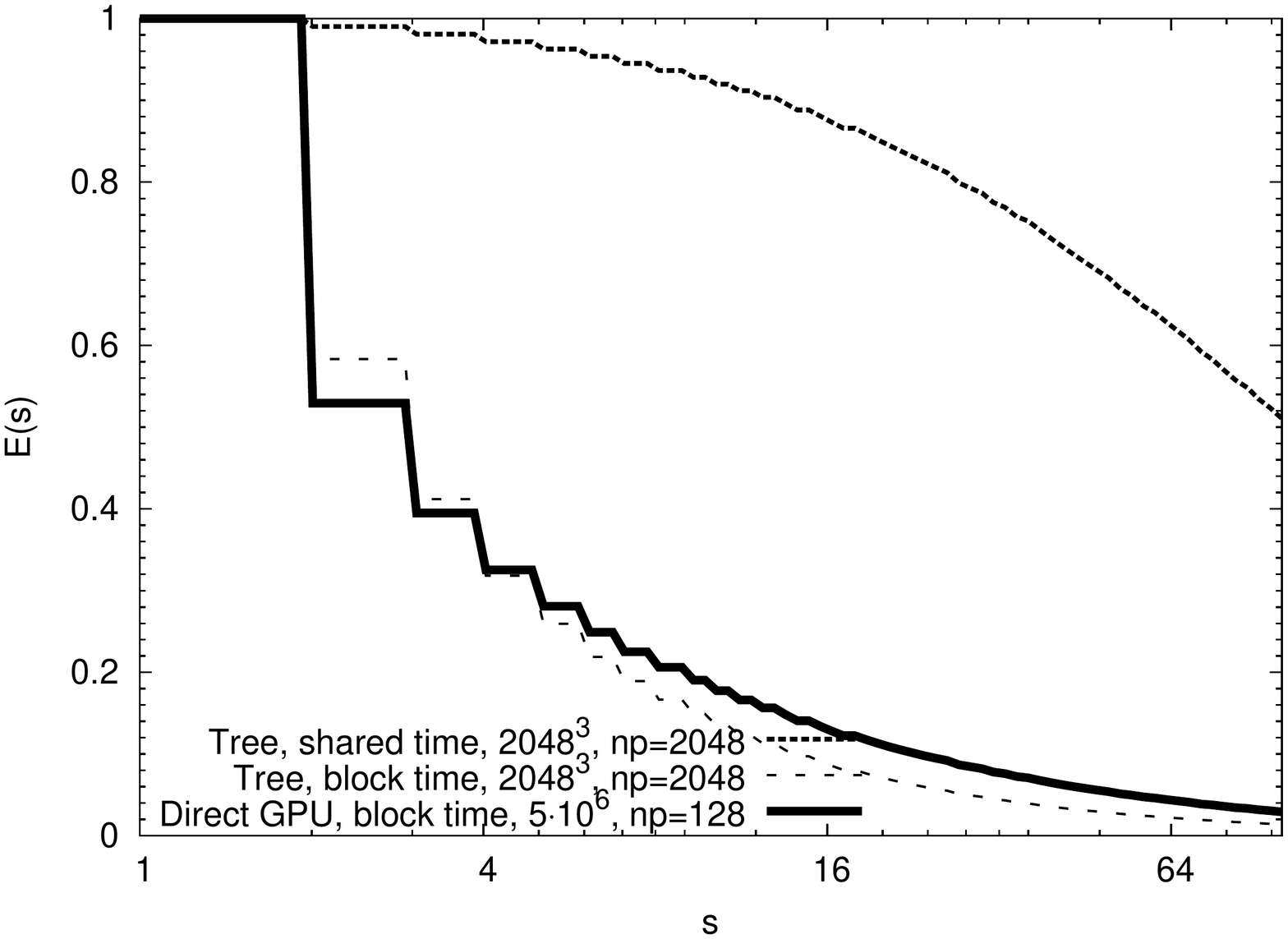}
  
  \caption{Predicted efficiency $E(s)$ of $N$-body simulations using the tree
  and direct method as a function of the number of sites $s$ in a global grid.
  The total number of processes is kept fixed for all simulations.}
  
  \label{Fig:Gefftree}
\end{figure}

We give the predicted efficiency of the three simulations over $s$
supercomputers relative to a simulation over one supercomputer, $E(s)$ in
Fig.~\ref{Fig:Gefftree}. The efficiency of these runs is mainly limited by the
latency-bound communication time. The tree code with shared time steps retains
a high efficiency for $s \aplt 16$, while the simulations with block time steps
are less efficient due to the larger number of communications. We predict a
slightly higher efficiency when $s \apgt 4$ for the direct-method simulation
than for the tree code simulation with block time steps. This is because the
lower $N$ results in a lower number of communication steps, and therefore in
less communication overhead.

\subsection{Bandwidth analysis for cosmological simulations}

\begin{figure}[h!]
  \includegraphics[width=0.9\columnwidth]{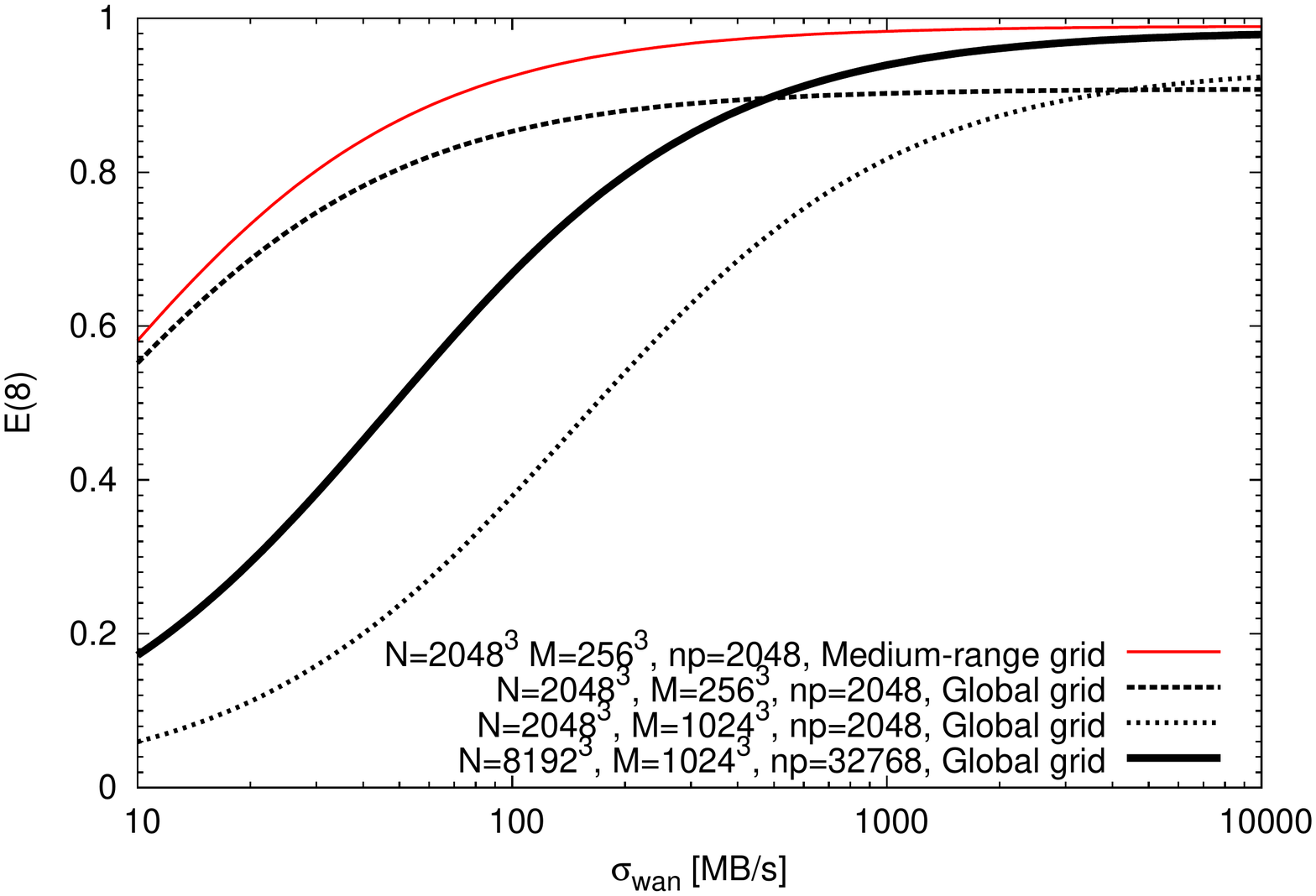}

  \caption{Predicted efficiency of four $N$-body simulations using the TreePM
  method over 8
  sites ($E(8)$) as a function of the wide area network throughput
  ($\sigma_{wan}$) in MB/s. Three simulations are run on a global grid. One
  simulation uses $N=2048^3$ particles and $M=256^3$ mesh cells (given by the
  black dashed line), one uses $N=2048^3$ and $M=1024^3$ (black dotted line)
  and one run uses $N=8192^3$ and $M=1024^3$ (thick black line). A simulation
  using $N=2048^3$ and $M=256^3$ modelled for a grid with 30 ms network latency
  between sites is given by the thin red line. Runs with $N=2048^3$ particles are
  predicted using a total of 2048 processes, runs with $N=8192^3$ particles are
  predicted using a total of 32768 processes.}

\label{Fig:DiscPerfBW}
\end{figure}

We have shown that our TreePM code scales well across up to $\sim 16$
sites if the ratio between the number of particles and the number of mesh cells
is sufficiently large. Here we examine the efficiency of four cosmological
simulations over 8 sites. The efficiency compared to a single site run is
predicted as a function of the available bandwidth. We have included three
predictions for a global grid with 0.3 s network latency as well as one
prediction for a simulation with $N=2048^3$ and $M=256^3$ over
a medium-range grid with 30 ms network latency. The details of these
simulations are given in Tab.~\ref{Tab:Nprobs}, and the results are shown in
Fig.~\ref{Fig:DiscPerfBW}.

We find that the efficiency of cosmological simulations across sites is heavily
dependent on the available bandwidth. The run with $N=2048^3$ and $M=256^3$ on
a global grid has $E(8) = 0.9$ (where $E(8)$ is the efficiency as defined in
Eq.~\ref{Eq:eff}) when the supercomputers are connected with a 100 MB/s
network. Using a network with a higher bandwidth has little effect on the
achieved efficiency, as the communication overhead is then dominated by network
latency. The effect of network latency is clearly visible when we look at the
prediction for the same simulation on a grid with a shorter baseline. When
using a grid with 30 ms network latency the simulation reaches $E(8) = 0.97$ if
the wide area network achieves a throughput of 1000 MB/s (which is possible
with a fine-tuned and dedicated 10 Gbps optical network). We predict an
efficiency $E(8) > 0.8$ for both simulations if the available network bandwidth
between sites is 50 MB/s.

The run with $N=2048^3$ and $M=1024^3$ is more communication intensive, and
requires a network throughput of at least 700 MB/s to achieve an efficiency of
0.8. The simulation using $N=8192^3$ and $M=1024^3$ runs more efficiently than
the simulation using $N=2048^3$ and $M=1024^3$ independent of the obtained
bandwidth. Although the exchanged data volume is larger in the run with
$N=8192^3$, this increased overhead is offset by the the higher force
calculation time per step. The large communication volume reduces the
efficiency considerably for low bandwidth values, and an average network
throughput of at least 150 MB/s is required to achieve an efficiency of
$E(8)=0.8$.

\section{Conclusion}

We have run a few dozen cosmological $N$-body simulations and analyzed the
scalability of our SUSHI integrator on a national distributed computer and
across a global network of supercomputers. Our results confirm that SUSHI is able to
efficiently perform simulations across supercomputers. We were able to run
a simulation using $1024^3$ particles across three supercomputers with $\sim
10\%$ communication overhead. The communication performance can be further
improved by tuning the optical networks.

Based on our model predictions we conclude that a long-term cosmological
simulation using $2048^3$ particles and $256^3$ mesh cells scales well over up
to $\sim 16$ sites, given that sufficient bandwidth is available and the number
of cores used per site is limited to $\sim 256$. We also predict that tree
codes with a shared time step scheme run efficiently across multiple
supercomputers, while tree codes with a block time step scheme do not.

Considerable effort is still required to obtain acceptable message passing
performance through a long distance optical network. This is due to three
reasons. First, it may take up to several months to arrange an intercontinental
light path. Second, optical networks are generally used for high volume data
streaming such as distributed visualization or bulk data transfer, and are
therefore not yet tuned to achieve optimal message passing performance. Third,
intercontinental networks traverse a large number of different institutes,
making it politically difficult for users to diagnose and adjust settings on
individual sections of the path. For our experiments we therefore chose to
optimize the wide area communications by tuning our application, rather than
requesting system-level modifications to the light path configuration.

The main challenges in running simulations across supercomputers are now
political, rather than technical. During the GBBP project, we were able to
overcome many of the political challenges in part due to good will of all
organizations involved and in part through sheer patience and perseverance.
However, orchestrating a reservation spanning across multiple supercomputers is
a major political undertaking. The use of a meta-scheduler and reservation
system for supercomputers and optical networks greatly reduces this 
overhead, and also improves the workload distribution between supercomputer
centers. Once the political barriers are overcome, we will be able to run
long lasting and large scale production simulations over a grid of supercomputers.

\section*{Acknowledgements}

We are grateful to Jeroen B\'edorf, Juha Fagerholm, Evghenii Gaburov, Kei
Hiraki, Wouter Huisman, Igor Idziejczak, Cees de Laat, Walter Lioen, Steve
McMillan, Petri Nikunen, Keigo Nitadori, Masafumi Oe, Ronald van der Pol, Gavin
Pringle, Steven Rieder, Huub Stoffers, Alan Verlo, Joni Virtanen and Seiichi
Yamamoto for their contributions to this work.
We also thank the network facilities of SURFnet, DEISA, IEEAF, WIDE, JGN2Plus, SINET3, 
Northwest Gigapop, the Global Lambda Integrated Facility (GLIF) GOLE of TransLight
Cisco on National LambdaRail, TransLight, StarLight, NetherLight, T-LEX,
Pacific and Atlantic Wave.
This research is supported by the Netherlands organization for Scientific
research (NWO) grant \#639.073.803, \#643.200.503 and \#643.000.803, the
Stichting Nationale Computerfaciliteiten (project \#SH-095-08), NAOJ, JGN2plus
(Project No. JGN2P-A20077), SURFNet (GigaPort project), the International
Information Science Foundation (IISF), the Netherlands Advanced School for
Astronomy (NOVA) and the Leids Kerkhoven-Bosscha fonds (LKBF). T.I. is
financially supported by Research Fellowship of the Japan Society for the
Promotion of Science (JSPS) for Young Scientists. This research is partially
supported by the Special Coordination Fund for Promoting Science and Technology
(GRAPE-DR project), Ministry of Education, Culture, Sports, Science and
Technology, Japan. 
We thank the DEISA Consortium (www.deisa.eu), co-funded through the
EU FP6 project RI-031513 and the FP7 project RI-222919, for support
within the DEISA Extreme Computing Initiative (GBBP project).

\bibliographystyle{unsrt}
\bibliography{Library}

\end{document}